\documentclass[aps,reprint,pra,superscriptaddress,floats,floatfix]{revtex4-2}

\usepackage{amsmath}
\usepackage{amsfonts}
\usepackage{amssymb}
\usepackage{graphicx}
\usepackage{color}
\usepackage[colorlinks=true,citecolor=blue,linkcolor=blue,urlcolor=blue]{hyperref}
\usepackage{times}
\usepackage{amstext}
\usepackage{latexsym}
\usepackage[running,mathlines]{lineno}
\usepackage{verbatim}
\usepackage{physics}
\usepackage{qcircuit}
\usepackage{multirow}
\usepackage{ulem}
\usepackage{soul}

\providecommand{\be}{\begin{equation}}
\providecommand{\ee}{\end{equation}}
\providecommand{\ba}{\begin{eqnarray}}
\providecommand{\ea}{\end{eqnarray}}
\newcommand{\mform}[1]{\begin{pmatrix} #1_{11} & \ldots & #1_{1N} \\
\vdots & \ddots & \vdots \\
#1_{N1} & \ldots & #1_{NN} \end{pmatrix}}
\newcommand{\vecform}[1]{\mqty(#1_1 \\ \vdots \\ #1_N)}

\newcommand{\im}{\mathrm{i}}
\newcommand{\ex}[1]{\mathrm{e}^{#1}}
\newcommand{\In}{\mathcal{I}}
\newcommand{\inv}[1]{{#1}^{-1}}
\newcommand{\PR}[1]{\ensuremath{\left[#1\right]}}
\newcommand{\PC}[1]{\ensuremath{\left(#1\right)}}

\newcommand{\M}{\mathcal{M}}
\newcommand{\N}{\mathcal{N}}
\newcommand{\Id}{\mathcal{I}}
\newcommand{\Hm}{\mathcal{H}}
\newcommand{\summ}{\sum_{m=0}^k}
\newcommand{\sumn}{\sum_{n=1}^k}

\newcommand{\xt}{\ket{x(t)}}
\newcommand{\xo}{\ket{x(0)}}
\newcommand{\eb}{\ket{b}}
\newcommand{\hc}[1]{{#1}^{\dagger}}

\let\oldref\ref
\renewcommand{\ref}[1]{(\oldref{#1})}

\begin{document}
\raggedbottom

\title{Detailed Account of Complexity for Implementation of Some Gate-Based Quantum Algorithms}

\author{Fernando R. Cardoso}
\affiliation{Departamento de F\'{i}sica, Universidade Federal de S\~{a}o Carlos, 13565-905 S\~{a}o Carlos, S\~{a}o Paulo, Brazil}

\author{Daniel Yoshio Akamatsu}
\affiliation{Departamento de F\'{i}sica, Universidade Federal de S\~{a}o Carlos, 13565-905 S\~{a}o Carlos, S\~{a}o Paulo, Brazil}

\author{Vivaldo Leiria Campo Junior}
\affiliation{Departamento de F\'{i}sica, Universidade Federal de S\~{a}o Carlos, 13565-905 S\~{a}o Carlos, S\~{a}o Paulo, Brazil}

\author{Eduardo I. Duzzioni} 
\affiliation{Departamento de F\' \i sica, Universidade Federal de Santa Catarina, Florian\'opolis, SC, 88040-900, Brazil}

\author{Alfredo Jaramillo Palma}
\affiliation{Escola de Engenharia de São Carlos, Universidade de São Paulo, 13566-590 S\~{a}o Carlos, S\~{a}o Paulo, Brazil}

\author{Celso J. Villas-Boas}
\affiliation{Departamento de F\'{i}sica, Universidade Federal de S\~{a}o Carlos, 13565-905 S\~{a}o Carlos, S\~{a}o Paulo, Brazil}

\begin{abstract}
In this review article, we are interested in the detailed analysis of complexity aspects of both time and space that arises from the implementation of a quantum algorithm on a quantum based hardware. In particular, some steps of the implementation, as state preparation and readout processes, in most of the cases can surpass the complexity aspects of the algorithm itself. We present the complexity involved in the full implementation of quantum algorithms for solving linear systems of equations and linear system of differential equations, from state preparation to the number of measurements needed to obtain good statistics from the final states of the quantum system, in order to assess the overall complexity of the processes.
\end{abstract}

\maketitle

\section{\label{sec:1}Introduction}

Quantum computing takes advantage of the unique properties of quantum mechanics, such as superposition and entanglement to carry out computational tasks in distinct ways than the classical computers do \cite{nielsen-chuang2002}. Since Richard Feynman's idealization that a quantum architecture would be a proper way to simulate actual quantum systems that occur in nature in the early $1980$'s \cite{feynman1982}, much attention has been given to the application of quantum systems for computational tasks. Among the greatest and most famous achievements of quantum information and quantum computation, one can cite superdense coding \cite{bennettsuperdense}, the BB-84 algorithm for quantum public key distribution of cryptography systems \cite{bb84}, Shor's integer factoring algorithm \cite{shor1994}, Grover's database search algorithm \cite{Grover1997}, alongside examples of no less importance or relevance. The advances have also reached important areas of mathematics and natural sciences in general, with quantum algorithms and circuit designing being developed to accomplish linear algebra tasks like eigen- \cite{Abrams1999,zhou2013} and singular value- \cite{rebentrost2018,gilyen2019} decompositions of matrices, finding solutions to linear systems of equations \cite{lloyd2009}, solving linear \cite{Berry2014,berry2017,Solano2020} and nonlinear \cite{leyton2008} differential equations, partial non-homogeneous linear differential equations \cite{Arrazola2019}, among other potential applications.   

There have been recent progress in the current era of Noisy Intermediate Scale Quantum (NISQ) devices, such as problems that cannot be solved by any classical shallow circuits in reasonable time, but turns out to be possible by shallow quantum circuits \cite{bravyi2020}, quantum supremacy using a superconducting quantum processor architecture achieved by Google team \cite{arute2019}, and also quantum advantages over classical computation using boson sampling \cite{zhong2020} and the simulation of quantum systems by means of quantum based architecture in D-Wave systems \cite{king2021}.

In general, the implementation of a quantum algorithm is based on many steps, that involve data pre-processing, preparation of input quantum states, the processing of the input information through quantum gates and operations done upon the system, measurement of the final state of the composite quantum system, and post-processing of the data collected by the measurement process. In the present work, we will not deal with the pre and post-processing steps, which are usually done by classical means. In most quantum algorithms, the quantum advantage over classical computation lies in the processing step, which takes advantage of the dimension of the Hilbert space of quantum systems and quantum parallelism to manipulate very large amounts of data, a task for which the present classical computers usually require exponential scaling resources, such as memory and state-of-the-art processors in supercomputer units. However, the preparation and measurement processes present in some quantum algorithms, which are essential for their proper implementations, are often neglected in their presentations, because of the intrinsic difficulties of these tasks.  

The main purpose of this work is to perform a detailed analysis of the computational complexity of quantum algorithms, considering all steps, from state preparation to readout processes. This work considers a scenario in which the rapid development of quantum computing has attracted the attention of people with different background, not only restricted to physicists or computer scientists from academia, but curious, investors, bankers, and entrepreneurs, which are delighted with the quantum speedups at first sight. Although quantum computing provides amazing results compared to classical computing, a suitable interpretation of the algorithmic complexity demands a proper analysis, taking into to account the conditions in which the results have been obtained and an accurate comparison with the corresponding classical task.

This work is organized as follows. In section \ref{sec:2} the complexity aspects of state preparation using different schemes are covered. Section \ref{sec:3} considers the complexity of quantum state tomography, with emphasis on the required number of measurements and repetitions of the execution of a quantum algorithm to achieve a desired accuracy in the results.  In section \ref{sec:5}, the overall complexity aspects for implementation are given, from state preparation to readout processes. Section \ref{sec:6} presents examples of quantum algorithms for solving linear systems of differential and algebraic equations, together with their complexity aspects. Finally, section \ref{sec:7} contains the conclusion of the work as well as some perspectives.  

\section{\label{sec:2} Complexity of Quantum State Preparation}

In many quantum algorithms implemented in the circuit model of Quantum Computation (QC) \cite{nielsen-chuang2002}, one is presented with the task of preparing quantum states as input for solving a given problem.
The input states constitute an important part in the process of implementation of a given algorithm for circuit gate-based quantum computing. As the final quantum state, encoding the solution of the problem, is directly linked to the input state through time-evolution, the complexity aspects of preparing the input state must be taken into account in a detailed resource analysis.
  
It is also worth noticing that there are some models of quantum computation in which the initial configuration of the constituting qubits must not be so specific for different tasks and problems. This is the case of Dissipative QC \cite{verstraete2009} and Measurement-Based QC \cite{briegel2009,van_den_nest2006}, which makes use of the so called Cluster States \cite{Dong2006,Raussendorf2003}, from which the final solution to the computational problem is given independently of a particular encoded configuration of initial conditions. 

To describe the encoding of input states properly, we must split the entire quantum system that constitutes a quantum computer into two parts: the ancilla qubits, which are used, for instance, to encode and control logical operations, and the work qubit system, that encodes the initial conditions of the problem to be solved and is submitted to the evolution process defined by the quantum algorithm. For instance, consider the processes to encode the initial conditions for a linear differential equation ~\cite{Solano2020} or for the HHL quantum linear problem ~\cite{lloyd2009} in the work system.  The goal of state preparation is to initialize the system in a $N$-dimensional specific quantum superposition that is suitable to the problem to be solved on a quantum computer. This task is often accomplished by subroutines that, in quantum algorithms, are usually referred to as system encoding. It is important to remark that there are different kinds of encoding, such as basis encoding and amplitude encoding: the former is used when one needs to manipulate real numbers arithmetically, and the latter when one takes advantage of the large size of the Hilbert space to encode data as probability amplitudes \cite{Leymann2020}.

In the case of basis encoding, the number of digits of the mantissa of the real number governs the precision of the approximation, and the exponent governs its range. Thus, when dealing with basis encoding, one must have an approximation with $n$ bits for the mantissa, $m$ bits for the exponent, and two additional bits to store the sign of both parts of the register, which results in a width of $(n+m+2)$ bits. To prepare such a state, a quantum-gate based circuit must be generated, in which the state of each component of the system is modified according to the desired encoding \footnote{It is worth noting that for binary encoding, the operations on each qubit can be done in parallel for a given string, resulting in a low depth circuit for this case of preparation}. E.g., binary encoded states are often used in D-Wave quantum processing units to solve optimization problems \cite{jiang2018,teplukhin2020}.

For amplitude encoding, the relevant information for computation is stored in the probability amplitudes of the quantum state, usually done starting from the $n$-qubit $\ket{0}^{\otimes n}$ by a transformation like

\begin{equation}\label{eq: gen-sup}
    \ket{0}^{\otimes n} \rightarrow \ket{\psi} = \sum_{i=0}^{N-1} c_i \ket{i}, 
\end{equation}
with $\sum_{i=0}^{N-1} \abs{c_i}^2 = 1$, and each $\ket{i}$ corresponding to a given state vector of the $N$-dimensional computational basis, with $N=2^n$. To address this task, one must be capable of preparing such superposition preserving coherence properties. The generic superposition can be prepared from the state $\ket{0}^{\otimes n}$ by applying single- and two-qubit operations directly to the system to be prepared, or by means of querying some device that contains previously stored information about the superposition \eqref{eq: gen-sup}.

In the most general case of a unitary transformation $\ket{\psi} = U \ket{0}^{\otimes n}$, the upper bound for the number of single and multi qubits operations scales as $O(N)$ \cite{shende2004,Leymann2020}. By manipulating the system directly, the qubit resource aspect for preparing such a superposition sums up to $O(n)$, as $\ket{\psi}$ represents a generic superposition of $n$ qubits. The state initialization can follow the procedure described in detail in \cite{long2001}, which makes use of standard single- and $\text{controlled}^k$-operations, which are operations controlled by $k$ qubits , acting on a single target. This method requires $O(N\log^2(N))$ single and two-qubit operations in total for executing a transformation like \eqref{eq: gen-sup} without the introduction of additional quantum bits. One should also take notice of the presence of $\text{controlled}^k$-operations, that can be further decomposed into $O(k^2)$ single and two-qubit quantum gates \cite{Barenco1995}. The particular structure of these controlled operations increases the depth of its action throughout the components of the quantum system \cite{long2001}. Soklakov and Schack present a quantum algorithm \cite{Soklakov2006} to prepare an arbitrary quantum register based on the Grover's search algorithm requiring resources that are polynomial in the number of qubits and additional gate operations $O(\textrm{polylog}(n))$.

\subsection{Quantum Database Creation and Search}

Employing calls on Random Access Memory (RAM) devices is an approach that aims to accomplish the task of preparation based on querying a database that contains the information of interest. In principle, it is possible to prepare a $N$-dimensional quantum superposition with a cost equivalent to $O(N)$ two-qubit operations per memory call with classical or quantum conventional RAM designs \cite{giovannetti2008}.

A quantum database consists in a set of state vectors which contains relevant information for quantum computation. For instance, suppose a set of $m$ vectors $S=\{x^1,x^2,\ldots,x^m\}$, each of them containing $k$ components. The quantum equivalent of this database is the quantum associative memory representation \cite{ventura2000} given by the uniform superposition of each state vector \cite{Leymann2020},

\begin{equation}
    \ket{S} = \frac{1}{\sqrt{m}} \sum_{i=1}^m \ket{x^i},
\end{equation}
for which the cost of the creation of $\ket{S}$ scales as $O(mp)$ \cite{ventura2000,Leymann2020}. Assuming that each $\ket{x^i}$ can be considered as a qubit system with dimension $k=N=2^n$, this would require $O(mN)$ steps, which is exponentially expensive in terms of the number of qubits of the referred system. Grover's quantum search algorithm is often used as subroutine for querying databases with complexity $O(\sqrt{m}\log(m))$ steps, while preparing and processing results of the query process would take $\Omega(m\log(N))$ steps \cite{Grover1997}.

It has been shown that the so called ``Bucket-Brigade" (BB) architecture for quantum RAM (qRAM) accomplishes the task of retrieving one memory cell previously prepared in a quantum superposition state coherently with $O(\log^2(N))$ \cite{giovannetti2008arch} steps. This architecture for qRAM is composed of a series of qutrits (three-level quantum systems described by the states $\ket{\textrm{wait}}$, $\ket{\textrm{left}}$, $\ket{\textrm{right}}$), which are used to guide the signals and retrieve the information stored in the memory cells. Thus, to completely construct the above superposition on a memory-call based scheme, one must be able to query the memory device at least $m$ times. This results in a query complexity that scales with $O(mN^2)$ and $O(m\log^2(N)$ two-qubit operations for conventional and BB-qRAM architectures, respectively, with data encoding in the memory cells amounting to $O(\sqrt{N})$ \cite{Leymann2020}. It is also remarkably noted that qRAM devices could constitute, in principle, a quantum system of $n$ qubits itself. By making such an observation, we point out that the number of qubits composing the full system (work+qRAM) also increases from $n$ to $2n$, which still scales as $O(\log (N))$.

The Flip-Flop qRAM (FF-qRAM) \cite{park2019} scheme has complexity aspects similar to BB-qRAM. Complexity amounts to $O(\log(N))$ additional qubits and $O(N\log(N))$ operations for reading unsorted data in previously stored in quantum memory cells to create a generic superposition as described by eq. \eqref{eq: gen-sup}. It has the advantage of not depending on proper routing algorithms, as it happens with the conventional and BB qRAM architectures \cite{giovannetti2008arch}, and is based on the quantum circuit computation model, what makes possible the application of quantum error-correction routines~\cite{park2019, Paetznick2013,anderson2014,Oconnor2014}.

\section{\label{sec:3}Complexity of Quantum State Tomography}

Quantum State Tomography (QST) is a procedure that aims for the complete reconstruction of an unknown density matrix $\rho$ \cite{nielsen-chuang2002}. Often, for information encoded in amplitudes or phases of a quantum state, after executing a quantum algorithm, one is presented with a density matrix whose elements ($\rho_{ij}$) codify the algorithm's output \cite{mohseni2008}. Information encoded in the complex amplitudes of a quantum state is not directly accessible through trivial means \cite{nielsen-chuang2002}. Thus, QST could represent a fundamental step in the knowledge of obtaining the full solution of a given problem. This consideration is important for a proper comparison between quantum and classical algorithms in which the quantum solution is a superposition state while the classical solution is a vector where all coefficients are known \cite{aaronson2015read}. At the same time quantum information can be storage in a Hilbert space whose dimension increases exponentially according to the number of qubits, to retrieve such information it is necessary to pay the price for that, which is also requires exponential steps. Alternatively, some global properties of the solution could be obtained by means of the expectation values of some observables, i.e., $\expval{O_k}=\Tr(O_k\rho)$\cite{Ekert2002}. This later approach usually conducts to quantum advantage in the processing time, however, it is not straightforward to get from average values of observables the desired quantities usually employed in practical applications of quantum computing. In this way, the impact of the QST complexity on the overall complexity of quantum algorithms must be carefully considered.

There are many quantum algorithms whose output state has coherence in a some basis. Our intention is not cover all articles in literature about this topic, but let us present some of them.  There are algorithms to solve partial differential equations \cite{Clader2013,cao2013quantum,montanaro2016quantum,costa2019quantum,Fillion2019,wang2020quantum}, linear differential equations \cite{Berry2014,berry2017,Solano2020}, nonlinear differential equations \cite{leyton2008quantum}, linear system of equations (also named quantum linear problem) \cite{lloyd2009,childs2017quantum,subacsi2019quantum}. In these examples, QST may be required depending on the level of detail expected to be known. For instance, the HHL algorithm \cite{lloyd2009} is particularly useful for obtaining a quantum state which results as the solution of a given system of coupled linear equations. It has been used as a subroutine for some of the above algorithms, yielding an exponential gain in the time processing of them. However, such exponential gain is achieved under particular hypothesis of easy state preparation (QRAM), sparseness of the matrix $A$ and if we are interested only in global properties of the solution encoded in the amplitudes of the work system state, which one can be extracted by expected values of the observables. Besides that, a detailed analysis of algorithms which use the finite element method and subsequently the HHL algorithm as subroutine to solve differential equations may result at most in polynomial gain \cite{montanaro2016quantum}. Another example shows that approaches based on HHL algorithm is never faster than the best classical algorithms \cite{linden2020quantum}.

There is a variety of QST processes and schemes available to accomplish the characterization task, such as Simple Quantum State Tomography (SQST) \cite{nielsen-chuang2002}, Ancilla Assisted Process Tomography~\footnote{Although reference \cite{Altepeter2003} discusses quantum process tomography, a QST procedure is needed in order to complete the protocol in SQPT and AAPT schemes, and an insight about the complexity of quantum state tomography can be obtained.} (AAPT) \cite{Altepeter2003}, QST via Linear Regression Estimation \cite{qi2013}, Compressed-Sensing QST \cite{Gross2010}, Principal Component Analysis \cite{lloyd2014}, and efficient process tomography \cite{cramer2010efficient}, each of these with particular complexity aspects, being suitable for specific problems. Their different computational complexities arise from taking advantage of particular characteristics of $\rho$. In particular, the basic idea of Compressed-Sensing is that a low-rank density matrix can be estimated with fewer copies of the state, as the sample complexity decreases with the rank $R$. Ref. \cite{flammia2012} introduces the matrix Dantzig selector and matrix Lasso estimators, with sample complexity for obtaining an estimate accurate within $\epsilon$ in trace distance scaling as $O(\frac{R^2N^2}{\epsilon^2}\log(N))$ for rank-$R$ states, requiring measuring of $O(RN\textrm{polylog}(N))$ Pauli expectation values.

In general, QST is based on the decomposition of the density matrix in a linear combination of basis operators. For a system of $n$ qubits, the dimension of the composed Hilbert space $\Hm = \otimes_{i=0}^{n-1} \Hm_i$ corresponds to $N=2^n$, and the reconstruction of a density matrix $\rho$ in such space requires $4^n - 1 = N^2 - 1$ basis operators \cite{nielsen-chuang2002}, which scales polynomially in the dimension $O(N^2)$, and exponentially in resources aspects $O(4^n)$. These exponential aspects of complexity are well known \cite{aaronson2007}. Besides the number of basis operators needed for characterization, it is important to remind that the reconstruction of $\rho$ is based on expected values of those basis operators. For instance, in the case of a single qubit, the set of $4^1 - 1 = 3$ basis operators needed for the proper quantum statistics could be based on the Pauli matrices $X$, $Y$, and $Z$, such that
\begin{equation}
    \rho = \frac{1}{2} \PR{\Tr(\rho) \Id + \Tr(\rho X)X + \Tr(\rho Y)Y + \Tr(\rho Z)Z}~,
\end{equation}
where $\Id$ is the identity operator. This statistical approach requires ensemble measurements of these observables, thus requiring a large number of copies~ $\rho$\cite{nielsen-chuang2002}. Besides these fundamental concepts, it has been shown that by using machine learning theory one could learn information about $\rho$ by a number of measurements that grow linearly with $n$ \cite{aaronson2007}. Ref. \cite{mohseni2008} gives a detailed description of the number of measurements and the scaling of the physical resources of the system. There are also models in which the QST problem is converted into a parameter estimation problem such as linear regression \cite{qi2013}, for which the computational complexity scales as $O(N^4)$.

The overall complexity aspects~\footnote{The overall complexity is defined as in \cite{mohseni2008}, given by the number of copies of $\rho$ times the number of gates per measurement.} yielded from SQST is $O(N^4\log(N))$, and the same relation holds for AAPT using Joint Separable Measurement (JSM) scheme. Both SQST and AAPT-JSM require only single body interactions \cite{mohseni2008}, while the Mutually Unbiased Bases (MUB) and the generalized POVM AAPT-schemes require many-body interactions, and their complexity scale as $O(N^2\log^2(N))  \PR{O(N^2\log^3(N))}$ and $O(N^4)$, respectively, under presence of nonlocal [local] two-body interactions. The Quantum Principal Component Analysis (QPCA) \cite{lloyd2014} focuses on reconstructing the eigenvectors of $\rho$ corresponding to the largest eigenvalues of the system, in a particular region of the space $\Hm$, in time $O(R \log(N))$, where $R$ stands for the rank of the corresponding density matrix. Compressed-Sensing, in contrast, reconstructs the full density matrix of the system in $O(RN\log^2(N))$ time steps~\cite{Gross2010}. The full density matrix reconstruction can also be realized with QPCA process, in a number of time steps that amounts to $O(RN\log(N))$\cite{lloyd2014}. 

In practice, all of the complexity aspects rising from measurement schemes used for obtaining prior information about the systems under consideration will increase the overall complexity of its implementation in quantum computing devices, which will be brought together in Sec. \ref{sec:5}.

\subsection{Pure state tomography}
\label{sec:tomography-positive-coefficients}

There exist certain procedures where one is not interested in the full description of the resulting state $\rho$ (e.g., some special cases of the algorithm in \ref{sec:6.1}). Instead, let us assume that the output of the algorithm is fully codified in the squares of the state's amplitudes, i.e., if $\ket{\Psi}=\sum_{m=1}^N \braket{m}{\Psi} \ket{m}$ is the output of the algorithm, then all one needs to know is each $|\braket{m}{\Psi}|^2$. More generally, one may be interested in knowing the square of the amplitudes associated to only a subspace of $\Hm$. An example of this is considered in \cite{Suzuki2020}, where it is assumed that the output of the algorithm can be written as
\begin{equation}
	\ket{\Psi} = \frac{1}{\mathcal{N}^2} \left( \ket{0} \ket{\Psi_0} +\ket{1}\ket{\Psi_1} \right)~,\label{eq:sol-compbase}
\end{equation}
where the first qubit is an auxiliary one, $\ket{\Psi_0}=\sum_{m=1}^N \alpha_m \ket{m}$ is the target state (written in terms of the computational base of the subsystem), $\ket{\Psi_1}$ is an arbitrary state, and $\mathcal{N}$ is a normalization constant that may depend on $N$. The probability of success $p$ corresponds to the probability of the auxiliar qubit to be found in the state $\ket{0}$, which may be computed as
\begin{equation}
	p=\frac{\braket{\Psi_0}}{\mathcal{N}^4}~.
\end{equation}
Moreover, the probability of the system to be found in the state $\ket{0}\ket{m}$ is $p_m=|\alpha_m|^2/\mathcal{N}^4$, here assumed to be non-null for every $m$. As explained in \cite{mohseni2008}, each $p_m$ is possible to be estimated by performing $M_m$ independent measurements, each measurement requiring one copy of $\ket{\Psi}$. After these trials, the probability $p_m$ is estimated as $\overline{p}_m=n_m/M_m$, where $n_m$ is the number of occurrences of $\ket{0}\ket{m}$. By statistical arguments, the authors also show that the number of trials necessary to estimate $p_m$ up to a relative precision $\Delta$ with probability $1-\epsilon$ \footnote{This exactly means that $|\overline{p}_m-p_m|/|p_m|\leq \Delta$ with probability $1-\epsilon$.}, denoted by $M_m(\Delta,\epsilon)$, is bounded as
\begin{equation}
	M_m\geq p_m^{-1}\,C(\vartriangle,\epsilon)~,\label{eq:Mm_pm}
\end{equation}
where $C(\vartriangle,\epsilon)\equiv \frac{3}{\Delta ^2}\log(1/\epsilon)$ does not depend on the system's size. Denote now the square of the normalized amplitude by $\beta_m^2\equiv |\alpha_m|^2/{\braket{\Psi_0}}$, then $p_m=|\alpha_m|^2/\mathcal{N}^4=p\,\beta_m^2$, and thus from \eqref{eq:Mm_pm} the behavior of $M_m$ can be determined from the behavior of $p$ and $\beta_m^2$ in terms of $N$. Now, assume that each $|\alpha_m|^2$ goes to 0 at the same rate as $N$ grows, i.e., $|\alpha_m|^2=O(N^{-r})$ for $r>0$ and all $m$. A particular case of the last occurs when the discrete probability distribution $\{\beta_m^2\}_m$ is fairly uniform, for which $r=1$. Therefore, since $\beta_m^2=|\alpha_m|^2/\sum_m|\alpha_m|^2$, one has that $\beta_m^2=1/N$ and from \eqref{eq:Mm_pm} the number of copies of $\ket{\Psi}$ necessary to determine each $p_m$, that can be taken as $M=\underset{m}{\max}\,{M_m}$, is such that 
\begin{equation}
M\geq O\left(p^{-1} \left(\underset{m}{\min}\, \beta_m^2\right)^{-1}\right)= O(p^{-1}N)~.    \label{eq:M-pinv}
\end{equation}
We conclude that if $p$ has a non-null minimum as a function of $N$, then the computational complexity of the tomography of all the $p_i$ is of order $N$. Otherwise, one needs to determine the asymptotic behavior of the success probability $p$ as $N$ grows (e.g., Section \ref{sec:6.1}).

\section{\label{sec:5}Overall Complexity of Implementation}
The overall complexity for implementation of a quantum algorithm accounts for all tasks that must be executed. It must take into account the total resource aspect, such as the number of work and ancilla qubits, that could eventually include qRAM systems, as well as the usual gate complexity aspect, brought together with the measurement complexity scheme. The last accounts for the number of copies and measurements done upon the final state in order to reconstruct its proper statistical averages and features.
We will start by the resource aspect. As discussed in sec. \ref{sec:2}, the preparation of a generic superposition can be done by manipulating the work system directly \cite{long2001} with no additional qubits needed. 
For the analysis of the corresponding overall gate complexity of an implementation, we need first to consider the amount of identical copies of $\rho$ needed for its proper reconstruction, given a determined scheme for the task \cite{mohseni2008}. Among the cases presented, the total number of copies of $\rho$ scales as $O(N^4)$ for SQPT, AAPT-JSM and Linear Regression tomography schemes, $O(N^2)$ for AAPT-MUB, and a single copy for AAPT-POVM scheme. This total number of copies will appear as a multiplying factor in the full complexity analysis, since all the operations in the implementation of the quantum algorithm, from preparation to readout, should be done this corresponding number of times.

\textbf{Preparation:} The overall complexity of the preparation step depends on whether it is implemented by operating directly on the work system or by queries made upon a previously prepared quantum RAM device \footnote{The complexity of preparing a quantum RAM device is beyond the scope of the present work.}. Without using ancillas in the preparation, its gate complexity amounts to $O(N \log^2(N))$ single- and two-qubit operations, and gate complexity for preparation using the Bucket-Brigade qRAM architecture \cite{giovannetti2008arch} presents $O(\log^2(N))$, as discussed in sec. \ref{sec:2}. The preparation implemented via FF-qRAM scheme is fully based on the quantum circuit computation model, without any routing algorithm to address the memory cells that must be queried throughout the transformation represented by \eqref{eq: gen-sup}. The number of gate operations in the FF-qRAM sums up to $O(\log(N))$ \cite{park2019}.

\textbf{Algorithm Evolution:} We define the expression \textit{algorithm evolution} to denote the process in which the previously prepared work system is evolved to its last configuration, which could represent, for instance, the solution of a system of linear equations \cite{lloyd2009}, a system of coupled differential equations \cite{Solano2020}, among other examples of possible applications for quantum computation. The quantum algorithm is composed by a sequence of defined steps and operations, which transforms the initial state under linear operations, that can be controlled by ancilla qubits that compose the full system under consideration. The evolution process will be denoted here as a linear map, represented by $\varepsilon$, as in ref. \cite{mohseni2008}. The gate and resource complexity of a given algorithm depends on the tasks that may be executed through its implementation, so different quantum algorithms have distinct complexity aspects, specially gate complexity. To represent generically the time complexity of the processing step of the algorithm, we will define a function $C(\varepsilon)$, which one excludes the steps of preparation and measurement of the quantum states.

\textbf{Readout:} The readout aspect must bring the analysis of the number of gates per measurement necessary to characterize a $N$-dimensional quantum system. For both SQTP and AAPT-JSM, $O(\log(N))$ single qubit operations must be implemented in order to reconstruct the density matrix. For AAPTs-MUB based schemes, one needs $O(\log^2(N))$ $\PR{O(\log^3(N))}$ single- and two-qubit gates, given that nonlocal [local] correlations occur in the system. The POVM scheme gate complexity scales as $O(N^4)$ \cite{mohseni2008}. There are, also, particular methods of reconstruction for $\rho$, such as QPCA \cite{lloyd2014} and Compressed-sensing, which are capable of reconstructing the density matrix with a number of gates up to $O(R \log(N))$ and $O(RN \log^2(N))$, where $R$ stands for the rank of the density matrix under reconstruction \cite{Gross2010}. For the application of those techniques, some knowledge of $\rho$ must be needed, such as the existence of larger eigenvalues in some regions of the composed Hilbert space \cite{lloyd2014} and sparsity of $\rho$. Since we assume that no prior information about $\rho$ is known, we do not discuss the overall complexity.

\textbf{Overall Complexity:} The overall gate complexity for implementation will now be classified according to each of the techniques discussed in detail in the previous sections. Considering a single shot of a quantum algorithm, represented by the action of the linear map $\varepsilon$, the gate complexity amounts to $O(N \log^2(N) + C(\varepsilon))$ in both SQTP and AAPT-JSM measurements schemes. For AAPT-MUB, the gate aspect sums up to $O(N\log^2(N) + C(\varepsilon) \PR{\log^3(N)})$, and to $O(\log^2(N) + C(\varepsilon) + N^4)$ for POVM scheme tomography \footnote{The presence of the term in brackets is necessary only if the system dealt with presents local interactions between the qubits. The nonlocal contributions results in a complexity aspect corresponding to $\log^2(N)$, which is already taken into account in the terms proportional to $\log^2(N)$ in the asymptotic notation.}. When FF-qRAM is used, the gate complexity per measure amounts to $O(\log(N) + C(\varepsilon))$ for SQTP/JSM measurements, $O(\log(N) + C(\varepsilon) + \log^2(N) [\log^3(N)])$ for MUB and $O(\log(N) + C(\varepsilon) + N^4)$ for POVM based schemes \cite{mohseni2008}. The overall aspects of gate complexity are represented in the table \ref{table: 1}. The combination of AAPT-MUB and POVM schemes for tomography with state preparation using the FF-qRAM \cite{park2019} provides, in terms of the dimension of the system, $N$, the lowest gate complexities, as it scales as $\approx O(N^2(\log(N)+C(\varepsilon)+\log^2(N)))$ and $O(N^4+\log(N)+C(\varepsilon))$, respectively.

\begin{table}
\caption{Gate Complexity Analysis for various schemes of preparation (DM - Direct Manipulation, BB - Bucket Brigade, FF - Flip-Flop) and readout (Measuring) procedures. The quantities in brackets are only taken into account if the system shows local interaction between qubits. The nonlocal interactions are already taken into account in the other terms, if the system shows such behavior. $C(\varepsilon)$ stands only for the time complexity of the processing stage of the quantum algorithm, represented via the linear map $\varepsilon$.}
\begin{tabular}{|c|c|c|}
\hline
\label{table: 1}
 & \multicolumn{2}{|c|}{Overall Gate Complexity} \\  
\hline
Measuring/Prep. & DM/BB-qRAM & FF-qRAM \\ 
\hline
\multirow{2}{4em}{SQTP/JSM} & $O(N^4(\log^2(N)$  & $O(N^4(\log(N)$ \\ 
	                            & $+C(\varepsilon)+\log(N)))$ & $+ C(\varepsilon)))$ \\
\hline
\multirow{3}{4em}{MUB} & $O(N^2(\log^2(N)$ & $O(N^2(\log(N)$  \\
                            & $+C(\varepsilon)+\log^2(N)$ & $+C(\varepsilon)+\log^2(N)$ \\ & $[\log^3(N)]))$ & $[\log^3(N)]))$ \\
\hline
\multirow{2}{4em}{POVM} & $O(\log^2(N)$ & $O(\log(N)+C(\varepsilon)$  \\
                            & $+C(\varepsilon)+N^4)$ & $+N^4)$ \\
\hline
\end{tabular}
\end{table}

\section{\label{sec:6}Examples - Complexity of Algorithms of Refs. \cite{Solano2020} and \cite{lloyd2009}}

\subsection{\label{sec:6.1} Complexity of algorithm in Ref. \cite{Solano2020}}

The complexity analysis of a quantum algorithm must take into account all the resources necessary for its implementation \cite{nielsen-chuang2002}, which involves both space and time resources. \cite{Solano2020} provides frameworks for solving both unitary and non-unitary evolutions of a linear differential equation (LDE), besides a general framework for solving any LDEs. This complexity analysis involves quantifying qubit resources and operations that must be applied on the composite quantum system to achieve the solution of the designated problem, which is described by the LDE
\begin{equation}\label{eq:5.1}
    \dv{\va{x}(t)}{t} = \M \va{x}(t) + \va{b},
\end{equation}
where $\va{x}(t)$ and $\va{b}(t)$ are $N$-dimensional vectors, and $\M$ a $N \times N$ matrix. This can be represented directly by
\begin{equation}\label{eq:5.2}
    \vecform{\dot{x}} = \mform{\M}\vecform{x} + \vecform{b}.
\end{equation}
 The solution to Eq. \eqref{eq:5.1} is given by $\va{x}(t) = \ex{\M t}\va{x}(0) + (\ex{\M t} - \In)\inv{\M} \va{b}$ and can be approximated by a Taylor expansion of order $k$ of the evolution operator $\ex{\M t}$ as
\begin{equation}\label{eq:5.3}
    \va{x}(t)\approx \summ \frac{\PC{\M t}^m}{m!} \va{x}(0) + \sumn \frac{\M^{n-1} t^n}{n!} \va{b}.
\end{equation}
To implement the quantum circuit, one must represent the vectors in the computational basis $\ket{x(0)} = \sum_j x_j(0)/\norm{x(0)} \ket{j}$, $\eb = \sum_j b_j/\norm{b}\ket{j}$, and matrix $\M$ as operators $A=\sum_{i,j} \M_{ij}/\norm{\M} \dyad{i}{j}$, where $\norm{\M}$ is the spectral norm of $\M$. Denoting $C_m=\norm{x(0)} \frac{(\norm{\M}t)^m}{m!}$, $D_n=\norm{b} \frac{(\norm{\M}t)^{n-1}t}{n!}$, and the operators $U_m=A^m$, the approximated solution is written (up-to normalization) as 
\begin{equation}
    \xt \approx \summ C_m U_m \xo + \sumn D_n U_{n-1} \eb~.
\end{equation}


For the sake of brevity, here the analysis is restricted to the case where the operator $A$ is unitary. Thus, all the powers of $A$ are unitary and the transformations performed on the work system can be implemented independently through controlled operations of ancilla qubits. 

\begin{figure}
\includegraphics[scale=0.75]{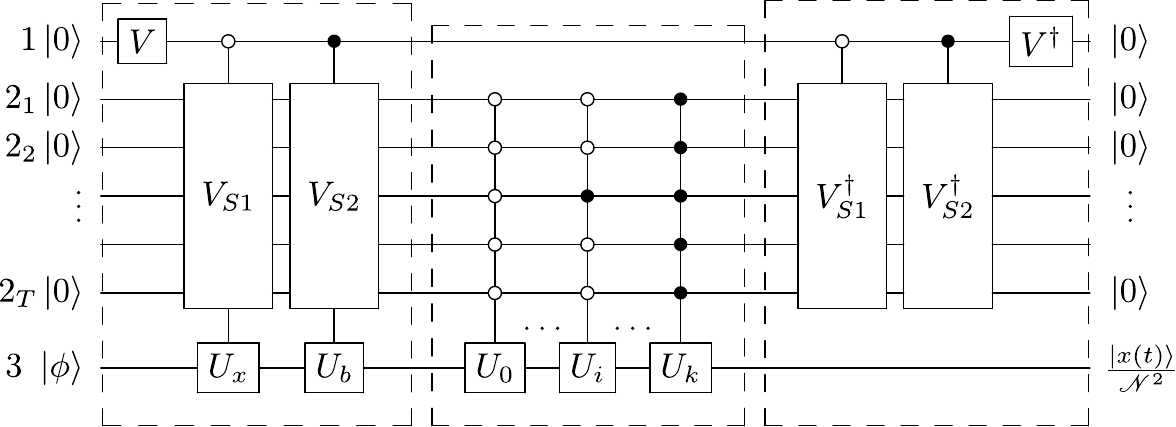}
\caption{\label{fig:A_unitary}Quantum circuit for solving LDEs related to a unitary operator $A$. Circuit representation of the unitary case. The operations are grouped by boxes corresponding to each of the information processing steps of the algorithm. From the left to the right, the boxes represent, respectively, the encoding, evolution and decoding steps. Figure adapted from Ref. \cite{Solano2020}.}
\end{figure}

The ancillary resources are represented by the number of ancilla qubits needed for the algorithm to work. For unitary $A$, the information about the Taylor expansion can be encoded into a Hilbert space of dimension $(k+1)$, which can be realized as a composite system of $\log(k+1)$ ancillary qubits. The work system employs $\log(N)$ qubits, and a register, consisting of a single qubit, to control the encoding and decoding operations. Thus, $1+\log(k+1)+\log(N)$ qubits are necessary to implement the unitary case.
The steps are divided as: encoding, evolution and decoding.

\textbf{Encoding}: The encoding step is represented by the grouped operations in the first box of figure \ref{fig:A_unitary}. The first gate acts on the first qubit register $1$, 

\begin{equation}
    V \ket{0} = \frac{1}{\N} \begin{pmatrix} C & D \\ D & -C \end{pmatrix} \ket{0}= \frac{1}{\N} \PC{C \ket{0}+ D \ket{1}},
\end{equation}
where $C=\sqrt{\sum_n C_n}$, $D=\sqrt{\sum_m D_m}$ and $\N^2 = C^2 + D^2$.

The following gates, $V_{S1}$ and $V_{S2}$ act on the second register, which consists of a set of $T=\log(k+1)$ qubits. The operations are controlled by the state of the first qubit register $1$ to prepare the states $\xo$ and $\eb$. The matrix form of $V_{S1}$ and $V_{S2}$ are

\begin{align}
    V_{S1} = \begin{pmatrix} \sqrt{C_0} & Q & \ldots & Q \\ \sqrt{C_1} & Q & \ldots & Q \\ \vdots & \vdots & \ddots & \vdots \\ \sqrt{C_k} & Q & \ldots & Q \end{pmatrix}, & \ V_{S2} = \begin{pmatrix} \sqrt{D_1} & Q & \ldots & Q \\ \vdots & \vdots & \ddots & \vdots \\ \sqrt{D_k} & Q & \ldots & Q \\ 0 & Q & \ldots & Q \end{pmatrix},
\end{align}
where the $Q$s are chosen arbitrarily to make the operators $V_{S1}$ and $V_{S2}$ unitary, and can be found by the Gram-Schmidt method \cite{Solano2020}. The gates $U_x$ and $U_b$ are applied to the work system through operations controlled by the first register $1$. The matrix forms of $U_x$ and $U_b$ depend on the boundary values of equation \eqref{eq:5.1}. The state of the full system after encoding is

\begin{equation}
\begin{split}
    \ket{\Psi} = \frac{1}{\N} \left[\ket{0} \PC{\summ \sqrt{C_m}\ket{m}} \xo \right. \\ + \left. \ket{1} \PC{\sumn \sqrt{D_n} \ket{n-1}} \eb  \right],
\end{split}
\end{equation}
where $\ket{n}$ is the state of all qubits in the second ancilla register according to the binary representation of $n$. The operators $V_{S1}$ and $V_{S2}$ can be decomposed in $O(k^2)$ elementary gates \cite{Solano2020}. To implement the preparation of $\xo$ and $\eb$, qRAM calls can be useful, such as FF-qRAM, with complexity amounting to $O(\log(N))$, as mentioned in Sec. \ref{sec:2}, resulting in $O(k^2 + \log(N))$ gate operations for preparation.

\textbf{Evolution}: The evolution step is represented by the controlled gates grouped in the second box of figure \ref{fig:A_unitary}. The gates $U_0$, $U_1$, $\ldots$, $U_k$, which represent the powers of the operator $A$ act on the work system, controlled by the states of the second ancilla register. The evolution is done by the operator

\begin{equation}
    U = \In_1 \otimes \dyad{0}_2 \otimes U_0 + \ldots + \In_1 \otimes \dyad{k}_2 \otimes U_k, 
\end{equation}
where $\In_1$ stands for the identity operator acting on the first qubit register, and $\dyad{k}_2$ stands for projections of states of the second ancilla register. After evolution, the state of the full system is left in the configuration

\begin{equation}
\begin{split}
    \ket{\Psi} = \frac{1}{\N} \left[ \ket{0} \PC{\summ \sqrt{C_m} \ket{m} U_m} \xo \right. \\ + \left. \ket{1} \PC{\sumn \sqrt{D_n} \ket{n-1} U_{n-1}} \eb \right].
\end{split}
\end{equation}
The controlled operations of the evolution step yields a gate complexity that scales as $O(k \log(k) \log(N))$ operations \cite{Solano2020}.


\textbf{Decoding}: The decoding step is represented by the grouped operations on the last box of figure \ref{fig:A_unitary}. The operations applied to the system in the steps of the encoding part are reversed. The gates $\hc{V}_{S1}$ and $\hc{V}_{S2}$ are applied on the second ancilla register $2$, controlled by the state of the first qubit register $1$, and the operation $\hc{V}$ on the first register. The complete expression for the full state of the system consists of a superposition of many ancillary states, in which the $Q$ coefficients will be present as amplitudes, 

\begin{equation}
\begin{split}
    \ket{\Psi_F} = \frac{1}{\N^2} \left[ \ket{0} \ket{0}^{\otimes T} \left( \summ C_m U_m \xo \right. \right. \\ + \left. \left. \sumn D_n U_{n-1} \eb \right) + \ket{\Phi} \right].
\end{split}
\end{equation}
The work system is left in the state proportional to the solution $\xt$ when the state of all ancilla qubits is $\ket{0}$, and $\ket{\Phi}$ being a state which is orthogonal to $\ket{0} \ket{0}^{\otimes T}$.

The gate complexity of the decoding step is similar to encoding, with reverse transformations $\hc{V}_{S1}$ and $\hc{V}_{S2}$ scaling as $O(k^2)$ elementary operations. By this detailed analysis of the algorithm, thus, it is concluded that the complexity of the algorithm scales as $O(k^2+\log(N)+k\log(k)\log(N))$ \cite{Solano2020} and, depending on the measurement scheme, has a multiplicative factor that depends on the dimension of the system, according to table \ref{table: 1}.

\subsection{Complexity dependence on the probability of success}

The probability of the first two registers to be in the state $\ket{0}\ket{0}^{\otimes T}$, named probability of success in Section \ref{sec:tomography-positive-coefficients}, can be written
\begin{equation}
p=\frac{\braket{x(t)}}{\N^4}~.
\end{equation}
Let us consider an example where the ordinary differential equation \eqref{eq:5.1} comes from discretising a 1-dimensional diffusion Partial Differential Equation by means of a Finite Difference Method. A typical case of this \cite{Leveque2007} results on the tridiagonal matrix
\begin{equation}
\M=N^2\begin{pmatrix}
	-2 &  1 &  &  &\\
	1 & -2 & 1 & &\\
	 & \ddots & \ddots & \ddots & \\
	 & & 1 & -2 & 1\\
	 & &  &1 & -2 
\end{pmatrix},
\end{equation}
which depends on the system's size $N$, and $\va{b}$ can be taken as the null vector. Assuming a non-null initial condition $\va{x}(0)$, it can be proved that both $\norm{\va{x}(0)}^2$ and $\norm{\va{x}(t)}^2$ are of $O(N)$ (for $t$ less than the stable time-step increment \cite{Leveque2007}), $\norm{\M}=O(N^2)$, $\N^4=O(N^{4k+1})$, where $k$ is the order of the Taylor's expansion, thus obtaining that $p=O(N^{-4k})$. Substituting these orders in \eqref{eq:M-pinv} one sees that the number of necessary copies of the algorithm's resulting state-vector goes as $M\geq O(N^{4k+1})$.

\subsection{\label{sec:6.2} Complexity of algorithm in Ref. \cite{lloyd2009}}

The HHL algorithm \cite{lloyd2009} provides a framework in which it is possible, using phase estimation and controlled rotation techniques \cite{chuang1997}, to prepare a quantum state which corresponds to the solution of a linear system of equations. The composed quantum system, after the implementation of all steps is left in a state which is proportional to the vector corresponding to the solution of the given linear system. Specifically, the problem under consideration is to solve 

\begin{equation}\label{eq: 6.1}
    A \vec{x} = \vec{b},
\end{equation}
for the unknown vector $\vec{x}$, where $A$ is a Hermitian \footnote{Non-Hermitian matrices are also tractable by the HHL algorithm, defining an operator $\Tilde{A}$, which is Hermitian and related to $A$.} $N \times N$ matrix.
Its solution corresponds to $\vec{x} = A^{-1}\vec{b}$, and the HHL algorithm is usually referred to as quantum linear problem. To perform this task, one must be able to represent $\vec{b}$ as a state vector $\ket{b}$, which is manipulated through the protocol to generate the state 

\begin{equation}
    \ket{x} = A^{-1}\ket{b},
\end{equation}
up to normalization.

For the implementation of the algorithm, three qubit registers are necessary: the first register consists of a single ancilla qubit, and it is used both to take part in controlled operations during the algorithm and to post-select the output state of the composite quantum system. The second register is composed by $m$ qubits, which are used during phase estimation to store the eigenvalues of $A$ to $m$-bit precision and to control operations on the other steps of the algorithm, and the third register, which consists of $\log(N)$ qubits, assuming that $\ket{b}$ is a $N$-dimensional state vector. Thus, the qubit requirements for implementation of the HHL subroutine has general complexity scaling as $O(1+m+\log(N))$. 

\begin{figure}
\includegraphics[scale=0.7]{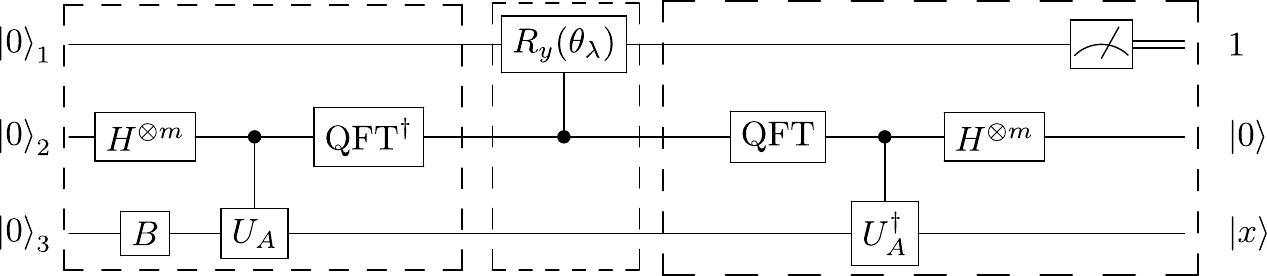}
\caption{\label{fig:circ-lloyd}Quantum circuit for the implementation of a linear system solver, according to the HHL algorithm. The different three stages of the algorithm are separated by the boxes indicated in the circuit.}
\end{figure}

The general procedure for implementation of the HHL subroutine algorithm is described as follows: in the first stage, the third register undergoes a procedure of Hamiltonian simulation, conditioned on the states of the second register, followed by phase estimation, to extract and store the eigenvalues of the operator $A$. In the second stage, controlled rotations are executed upon the first qubit register, controlled by the second register, to extract the inverse of the eigenvalues, $\lambda_j^{-1}$, and store the information into a superposition of entangled states between the registers. The third stage is where the phase estimation procedures of the first stage are reversed, leading the state of the second register to the initial configuration, and the first and third register in a superposition of entangled states in which, by measuring the state of the first, the state of the whole system is post-selected to 

\begin{equation}
    \ket{\Psi_{\text{f}}} \propto \ket{1}\ket{0}^{\otimes m}\ket{x},
\end{equation}
where $\ket{x}=\sum_j \beta_j/\lambda_j \ket{j}$ represents the solution of the given linear system. The corresponding generic circuit representation of the algorithm is given in figure (\ref{fig:circ-lloyd}). A detailed description of the stages of implementation is given in the following.

\textbf{First Stage:} In the first stage of the algorithm, the state $\ket{b}$ is prepared on the third register. For this task, it is assumed that there exists a unitary operation $B$, acting on $\log(N)$ qubits, which can transform an initial quantum state in the state $\ket{b}$. For simplicity, and without loss of generality, this initial state will be assumed to be $\ket{0}^{\otimes n}$, and the transformation $B$ executes the task $\ket{0}^{\otimes n} \rightarrow \ket{b} = \sum_j b_j \ket{j}$ in $T_{\text{B}}$ steps, being $\ket{j}$ the $N$-dimensional computational basis state. The second register is also transformed by the action of the $m$-qubit Hadamard gate, thus leaving the composite system in the configuration

\begin{equation}
    \frac{1}{2^{m/2}} \ket{0}_1 \sum_{k=0}^{M-1} \ket{k}_2 \ket{b}_3,
\end{equation}
with $M=2^m$. It is worth noting that these $T_B$ steps used to prepare the initial state $\ket{b}$ of the work system can be executed by any of the preparation processes described before, either by direct manipulation of the work system by unitary operations, or by direct calls on a previously established qRAM structure, resulting in $O(\log(N))$ steps.

The next operation is the conditioned Hamiltonian simulation executed upon the third register. This operation is controlled by states of the qubits from the second register, and is represented by $U_{\text{A}} = \sum_k \dyad{k}_2 \otimes \ex{\im A k t_0/M}$. Assuming the operator $A$ can be represented in the diagonal form $A = \sum_j \lambda_j \dyad{j}_3$, where $\lambda_j$ represents the eigenvalues of $A$, the action of $U_{\text{A}}$ leads to the composite state

\begin{equation}\label{eq: l-2}
    \frac{1}{2^{m/2}} \ket{0}_1 \sum_{k=0}^{M-1} \sum_{j=0}^{N-1} \ex{2 \pi \im \Tilde{\lambda}_j k/M} \beta_j \ket{k}_2 \ket{j}_3, 
\end{equation}
where $\Tilde{\lambda}_j$ represents the $m$-bit representation of the $j^{\text{th}}$ eigenvalue of the operator $A$. The sum over $k$ in eq. \eqref{eq: l-2} is the definition of a Fourier transformed $m$-qubit state, which can be expressed, equivalently, in the product representation \cite{nielsen-chuang2002}. Thus, applying the inverse Fourier transform operation $\hc{\text{QFT}}$ upon the second register results in

\begin{equation}
    \ket{0}_1 \sum_{j=0}^{N-1} \beta_j \ket{\Tilde{\lambda}_j}_2 \ket{j}_3,
\end{equation}
where the eigenvalues are now encoded in the $m$-bit strings of the states of the second register. These will be latter used to control operations on the first register in the second stage of the algorithm. The Hamiltonian simulation $U_A$ can be done in $O(s^2 t \log(N))$ steps assuming the matrix $A$ to be $s$-sparse \cite{lloyd2009}. The quantum Fourier transforms $\hc{\mathrm{QFT}}$ and $\mathrm{QFT}$ can be implemented in the $m$-qubit register $\Theta(m^2)$ steps, and the Hadamard operations $H^{\otimes m}$ consists in $O(m)$ single qubit rotations. This results in complexity aspects $O(m^2+s^2t\log(N))$

\textbf{Second stage:} In this stage, one must execute conditional rotations on the first register. These operations are controlled by the states of the second register, namely at this point, $\ket{\Tilde{\lambda}_j}$. The aim of these operations is to extract $1/\Tilde{\lambda}_j$, for each eigenvalue $\lambda_j$ and corresponding eigenvector $\ket{j}$. This operation consists on a controlled-$m$ qubit rotation on the first register, which is represented in the circuit of figure (\ref{fig:circ-lloyd}) by $R_y(\theta_{\lambda})$. This operation has the matrix form 

\begin{equation}
    R_y(\theta_j) = \mqty(\cos(\theta_j) & -\sin(\theta_j) \\
                          \sin(\theta_j) &  \cos(\theta_j)),
\end{equation}
with $\theta_j = \arcsin{\left(C/\Tilde{\lambda}_j\right)}$ where $C$ is a constant. The controlled operation is then represented by $\sum_j \dyad{\Tilde{\lambda}_j}_2 \otimes R_y(\theta_j)$. At this stage, $m$-controlled operations are executed on the first register, each corresponding to a rotation depending on the state of the second register. With these operations, the state is transformed into

\begin{equation}
\begin{split}
    \sum_{j=0}^{N-1} \left[\cos(\theta_j) \ket{0}_1 + \sin(\theta_j) \ket{1}_1\right] \beta_j \ket{\Tilde{\lambda}_j}_2 \ket{u_j}_3 \\
    =\sum_{j=0}^{N-1} \left[ \sqrt{1-\frac{C^2}{\Tilde{\lambda}^2_j}}\ket{0}_1 + \frac{C}{\Tilde{\lambda}_j}\ket{1}_1 \right] \beta_j \ket{\Tilde{\lambda}_j}_2 \ket{u_j}_3.
\end{split}
\end{equation}

\textbf{Third Stage:} In the third and last stage of the HHL algorithm, the operations executed over the system in the first stage, as the Quantum Fourier transform and Hamiltonian evolution are inverted, and a measurement on the first qubit register is done for postselection. If the outcome of this measurement is $1$, then the algorithm had succeeded in preparing a quantum state proportional to $\ket{x}$. 

The first operation of the third stage is the Quantum Fourier transform on the second register. This has the effect of leaving the second register in the configuration 

\begin{equation}
    \ket{\Tilde{\lambda}_j}_2 \rightarrow \frac{1}{2^{m/2}} \sum_{k=0}^{M-1} \ex{\frac{2\pi \im \Tilde{\lambda}_j k}{2^m}} \ket{k}_2.
\end{equation}

With the state of the second register again being represented by a superposition of the $\ket{k}_2$ states, the next step is to apply reverse controlled operation, $\hc{U}_A$, to the third register. This operation has the form $\hc{U}_A = \sum_{k'} \dyad{k'}_2 \ex{-\frac{2 \pi \im k' A}{2^m}}$. With these controlled operations, the relative exponential factors are eliminated from the superposition, and the state of the composite system is represented by

\begin{equation}
    \frac{1}{2^{m/2}} \sum_{j=0}^{N-1} \sum_{k=0}^{M-1} \beta_j  \left[ \sqrt{1-\frac{C^2}{\Tilde{\lambda}^2_j}}\ket{0}_1 + \frac{C}{\Tilde{\lambda}_j}\ket{1}_1 \right] \ket{k}_2 \ket{u_j}_3. 
\end{equation}
The second register is then submited to the $m$-qubit Hadamard transform, as done in the first stage. This operation has the effect of bringing back the state $1/2^{m/2} \sum_k \ket{k}$ to the configuration $\ket{0}^{\otimes m}$, and thus leaving the composite system in the state

\begin{equation}
    \sum_{j=0}^{N-1} \left[ \sqrt{1-\frac{C^2}{\Tilde{\lambda}^2_j}}\ket{0}_1 + \frac{C}{\Tilde{\lambda}_j}\ket{1}_1 \right] \ket{0}_2^{\otimes m} \beta_j \ket{u_j}_3.
\end{equation}

Then, the measurement is made upon the first register, aiming to post-select the quantum state in the third register which is proportional to the vector of solutions, $\ket{x}$. If the outcome $1$ is obtained, the state 

\begin{equation}
    C \sum_{j=0}^{N-1} \frac{\beta_j}{\Tilde{\lambda}_j} \ket{0}_2^{\otimes m} \ket{u_j}_3
\end{equation}
is heralded, with probability $~ \sum_j \abs{\frac{C \beta_j}{\Tilde{\lambda}_j}}^2$ \cite{lloyd2009}, that does not affect the computational complexity of the problem. An important observation is that the fidelity of the state is not with respect to $\ket{x}$ itself after post-selection, but to $\ket{0}^{\otimes m}_2 \ket{x}_3$. The complexity aspects for the third stage are similar to the first, but without Hamiltonian simulation. 

The overall aspects of gate complexity of the HHL algorithm are given as $O(m^2+s^2 t \log(N) + \log(N))$, with almost all operations concentrated in the first and third steps of the quantum algorithm, consisting in creating the initial superposition $\ket{b}$, Hamiltonian simulation of the operator $A$ and the $\mathrm{QFT}$ procedures. It is worth noting that the expectation values to extract global properties of the solution can imply in additional complexity aspects, and depends strongly on the probability amplitudes of the final state and the schemes used for measurement.

\section{\label{sec:7}Conclusion}

We present in this work a theoretic overview of the total complexity for implementation of some gate-based quantum algorithms which involve the codification of the system parameter in the initial state of the work/register qubits and the solution to the given problem is also codified in the final state of the qubtis, represented by a generic linear map $\varepsilon$. It is important to notice that algorithms that depend on the preparation of input states as superpositions of the basis states have at least $O(N\log^2(N))$ gate operations based on direct manipulation of the work qubits. Once a FF-qRAM device is available, this complexity can be reduced to $O(\log(N))$, assuming that the state is previously prepared in such a device. Concerning the output states, the analysis is made considering a fairly uniform probability distribution, as the precise estimation of the number of samples needed for tomography depends upon the probability distribution of states on the Hilbert space. In this case, if the desired result is encoded in a single amplitude of a given state, the number of required ensemble copies scales as $O(N)$ in the best case scenario. For algorithms depending upon the preparation of a superposition state, and the solution is encoded in the final superposition state of the work qubits, the overall complexity will be at least $O(N\log(N)C(\varepsilon))$, $C(\varepsilon)$ being the time complexity of the processing stage of the quantum algorithm. We point out that this complexity overview also depends on the architecture of the quantum hardware in which the algorithm should be implemented, and the availability of basic quantum gates for proper decomposition of all operations needed in the process of implementation.

\begin{acknowledgments}
This work was supported by the Coordenação de Aperfeiçoamento de Pessoal de  Nível Superior (CAPES)
- Finance Code 001, and through the CAPES/STINT project, grant No.
88881.304807/2018-01. C.J.V.-B. is also grateful for the
support by the São Paulo Research Foundation (FAPESP)
Grant No. 2019/11999-5, and the
National Council for Scientific and Technological Development (CNPq) Grant No. 307077/2018-7. This work is also part of the Brazilian National Institute of Science and Technology for Quantum Information
(INCT-IQ/CNPq) Grant No. 465469/2014-0.
\end{acknowledgments}

\bibliography{refs1}

\begin{thebibliography}{69}%
\makeatletter
\providecommand \@ifxundefined [1]{%
 \@ifx{#1\undefined}
}%
\providecommand \@ifnum [1]{%
 \ifnum #1\expandafter \@firstoftwo
 \else \expandafter \@secondoftwo
 \fi
}%
\providecommand \@ifx [1]{%
 \ifx #1\expandafter \@firstoftwo
 \else \expandafter \@secondoftwo
 \fi
}%
\providecommand \natexlab [1]{#1}%
\providecommand \enquote  [1]{``#1''}%
\providecommand \bibnamefont  [1]{#1}%
\providecommand \bibfnamefont [1]{#1}%
\providecommand \citenamefont [1]{#1}%
\providecommand \href@noop [0]{\@secondoftwo}%
\providecommand \href [0]{\begingroup \@sanitize@url \@href}%
\providecommand \@href[1]{\@@startlink{#1}\@@href}%
\providecommand \@@href[1]{\endgroup#1\@@endlink}%
\providecommand \@sanitize@url [0]{\catcode `\\12\catcode `\$12\catcode
  `\&12\catcode `\#12\catcode `\^12\catcode `\_12\catcode `\%12\relax}%
\providecommand \@@startlink[1]{}%
\providecommand \@@endlink[0]{}%
\providecommand \url  [0]{\begingroup\@sanitize@url \@url }%
\providecommand \@url [1]{\endgroup\@href {#1}{\urlprefix }}%
\providecommand \urlprefix  [0]{URL }%
\providecommand \Eprint [0]{\href }%
\providecommand \doibase [0]{https://doi.org/}%
\providecommand \selectlanguage [0]{\@gobble}%
\providecommand \bibinfo  [0]{\@secondoftwo}%
\providecommand \bibfield  [0]{\@secondoftwo}%
\providecommand \translation [1]{[#1]}%
\providecommand \BibitemOpen [0]{}%
\providecommand \bibitemStop [0]{}%
\providecommand \bibitemNoStop [0]{.\EOS\space}%
\providecommand \EOS [0]{\spacefactor3000\relax}%
\providecommand \BibitemShut  [1]{\csname bibitem#1\endcsname}%
\let\auto@bib@innerbib\@empty
\bibitem [{\citenamefont {Nielsen}\ and\ \citenamefont
  {Chuang}(2002)}]{nielsen-chuang2002}%
  \BibitemOpen
  \bibfield  {author} {\bibinfo {author} {\bibfnamefont {M.~A.}\ \bibnamefont
  {Nielsen}}\ and\ \bibinfo {author} {\bibfnamefont {I.}~\bibnamefont
  {Chuang}},\ }\href@noop {} {\bibinfo {title} {Quantum computation and quantum
  information}} (\bibinfo {year} {2002})\BibitemShut {NoStop}%
\bibitem [{\citenamefont {Feynman}(1982)}]{feynman1982}%
  \BibitemOpen
  \bibfield  {author} {\bibinfo {author} {\bibfnamefont {R.~P.}\ \bibnamefont
  {Feynman}},\ }\bibfield  {title} {\bibinfo {title} {Simulating physics with
  computers},\ }\href {https://doi.org/10.1007/BF02650179} {\bibfield
  {journal} {\bibinfo  {journal} {Int. J. Theor. Phys}\ }\textbf {\bibinfo
  {volume} {21}} (\bibinfo {year} {1982})}\BibitemShut {NoStop}%
\bibitem [{\citenamefont {Bennett}\ and\ \citenamefont
  {Wiesner}(1992)}]{bennettsuperdense}%
  \BibitemOpen
  \bibfield  {author} {\bibinfo {author} {\bibfnamefont {C.~H.}\ \bibnamefont
  {Bennett}}\ and\ \bibinfo {author} {\bibfnamefont {S.~J.}\ \bibnamefont
  {Wiesner}},\ }\bibfield  {title} {\bibinfo {title} {Communication via one-
  and two-particle operators on einstein-podolsky-rosen states},\ }\href
  {https://doi.org/10.1103/PhysRevLett.69.2881} {\bibfield  {journal} {\bibinfo
   {journal} {Phys. Rev. Lett.}\ }\textbf {\bibinfo {volume} {69}},\ \bibinfo
  {pages} {2881} (\bibinfo {year} {1992})}\BibitemShut {NoStop}%
\bibitem [{\citenamefont {Bennett}\ and\ \citenamefont
  {Brassard}(1984)}]{bb84}%
  \BibitemOpen
  \bibfield  {author} {\bibinfo {author} {\bibfnamefont {C.~H.}\ \bibnamefont
  {Bennett}}\ and\ \bibinfo {author} {\bibfnamefont {G.}~\bibnamefont
  {Brassard}},\ }\bibfield  {title} {\bibinfo {title} {Quantum cryptography:
  Public key distribution and coin tossing},\ }\href@noop {} {\bibfield
  {journal} {\bibinfo  {journal} {arXiv preprint arXiv:2003.06557}\ } (\bibinfo
  {year} {1984})}\BibitemShut {NoStop}%
\bibitem [{\citenamefont {Shor}(1994)}]{shor1994}%
  \BibitemOpen
  \bibfield  {author} {\bibinfo {author} {\bibfnamefont {P.~W.}\ \bibnamefont
  {Shor}},\ }\bibfield  {title} {\bibinfo {title} {Algorithms for quantum
  computation: discrete logarithms and factoring}\ }(\bibinfo {organization}
  {Ieee},\ \bibinfo {year} {1994})\ pp.\ \bibinfo {pages}
  {124--134}\BibitemShut {NoStop}%
\bibitem [{\citenamefont {Grover}(1997)}]{Grover1997}%
  \BibitemOpen
  \bibfield  {author} {\bibinfo {author} {\bibfnamefont {L.~K.}\ \bibnamefont
  {Grover}},\ }\bibfield  {title} {\bibinfo {title} {Quantum computers can
  search arbitrarily large databases by a single query},\ }\href
  {https://doi.org/10.1103/PhysRevLett.79.4709} {\bibfield  {journal} {\bibinfo
   {journal} {Phys. Rev. Lett.}\ }\textbf {\bibinfo {volume} {79}},\ \bibinfo
  {pages} {4709} (\bibinfo {year} {1997})}\BibitemShut {NoStop}%
\bibitem [{\citenamefont {Abrams}\ and\ \citenamefont
  {Lloyd}(1999)}]{Abrams1999}%
  \BibitemOpen
  \bibfield  {author} {\bibinfo {author} {\bibfnamefont {D.~S.}\ \bibnamefont
  {Abrams}}\ and\ \bibinfo {author} {\bibfnamefont {S.}~\bibnamefont {Lloyd}},\
  }\bibfield  {title} {\bibinfo {title} {Quantum algorithm providing
  exponential speed increase for finding eigenvalues and eigenvectors},\ }\href
  {https://doi.org/10.1103/PhysRevLett.83.5162} {\bibfield  {journal} {\bibinfo
   {journal} {Phys. Rev. Lett.}\ }\textbf {\bibinfo {volume} {83}},\ \bibinfo
  {pages} {5162} (\bibinfo {year} {1999})}\BibitemShut {NoStop}%
\bibitem [{\citenamefont {Zhou}\ \emph {et~al.}(2013)\citenamefont {Zhou},
  \citenamefont {Kalasuwan}, \citenamefont {Ralph},\ and\ \citenamefont
  {O'brien}}]{zhou2013}%
  \BibitemOpen
  \bibfield  {author} {\bibinfo {author} {\bibfnamefont {X.-Q.}\ \bibnamefont
  {Zhou}}, \bibinfo {author} {\bibfnamefont {P.}~\bibnamefont {Kalasuwan}},
  \bibinfo {author} {\bibfnamefont {T.~C.}\ \bibnamefont {Ralph}},\ and\
  \bibinfo {author} {\bibfnamefont {J.~L.}\ \bibnamefont {O'brien}},\
  }\bibfield  {title} {\bibinfo {title} {Calculating unknown eigenvalues with a
  quantum algorithm},\ }\href {https://doi.org/10.1038/nphoton.2012.360}
  {\bibfield  {journal} {\bibinfo  {journal} {Nature photonics}\ }\textbf
  {\bibinfo {volume} {7}},\ \bibinfo {pages} {223} (\bibinfo {year}
  {2013})}\BibitemShut {NoStop}%
\bibitem [{\citenamefont {Rebentrost}\ \emph {et~al.}(2018)\citenamefont
  {Rebentrost}, \citenamefont {Steffens}, \citenamefont {Marvian},\ and\
  \citenamefont {Lloyd}}]{rebentrost2018}%
  \BibitemOpen
  \bibfield  {author} {\bibinfo {author} {\bibfnamefont {P.}~\bibnamefont
  {Rebentrost}}, \bibinfo {author} {\bibfnamefont {A.}~\bibnamefont
  {Steffens}}, \bibinfo {author} {\bibfnamefont {I.}~\bibnamefont {Marvian}},\
  and\ \bibinfo {author} {\bibfnamefont {S.}~\bibnamefont {Lloyd}},\ }\bibfield
   {title} {\bibinfo {title} {Quantum singular-value decomposition of nonsparse
  low-rank matrices},\ }\href {https://doi.org/10.1103/PhysRevA.97.012327}
  {\bibfield  {journal} {\bibinfo  {journal} {Phys. Rev. A}\ }\textbf {\bibinfo
  {volume} {97}},\ \bibinfo {pages} {012327} (\bibinfo {year}
  {2018})}\BibitemShut {NoStop}%
\bibitem [{\citenamefont {Gily{\'e}n}\ \emph {et~al.}(2019)\citenamefont
  {Gily{\'e}n}, \citenamefont {Su}, \citenamefont {Low},\ and\ \citenamefont
  {Wiebe}}]{gilyen2019}%
  \BibitemOpen
  \bibfield  {author} {\bibinfo {author} {\bibfnamefont {A.}~\bibnamefont
  {Gily{\'e}n}}, \bibinfo {author} {\bibfnamefont {Y.}~\bibnamefont {Su}},
  \bibinfo {author} {\bibfnamefont {G.~H.}\ \bibnamefont {Low}},\ and\ \bibinfo
  {author} {\bibfnamefont {N.}~\bibnamefont {Wiebe}},\ }\bibfield  {title}
  {\bibinfo {title} {Quantum singular value transformation and beyond:
  exponential improvements for quantum matrix arithmetics}\ }(\bibinfo {year}
  {2019})\ pp.\ \bibinfo {pages} {193--204}\BibitemShut {NoStop}%
\bibitem [{\citenamefont {Harrow}\ \emph {et~al.}(2009)\citenamefont {Harrow},
  \citenamefont {Hassidim},\ and\ \citenamefont {Lloyd}}]{lloyd2009}%
  \BibitemOpen
  \bibfield  {author} {\bibinfo {author} {\bibfnamefont {A.~W.}\ \bibnamefont
  {Harrow}}, \bibinfo {author} {\bibfnamefont {A.}~\bibnamefont {Hassidim}},\
  and\ \bibinfo {author} {\bibfnamefont {S.}~\bibnamefont {Lloyd}},\ }\bibfield
   {title} {\bibinfo {title} {Quantum algorithm for linear systems of
  equations},\ }\href {https://doi.org/10.1103/PhysRevLett.103.150502}
  {\bibfield  {journal} {\bibinfo  {journal} {Phys. Rev. Lett.}\ }\textbf
  {\bibinfo {volume} {103}},\ \bibinfo {pages} {150502} (\bibinfo {year}
  {2009})}\BibitemShut {NoStop}%
\bibitem [{\citenamefont {Berry}(2014)}]{Berry2014}%
  \BibitemOpen
  \bibfield  {author} {\bibinfo {author} {\bibfnamefont {D.~W.}\ \bibnamefont
  {Berry}},\ }\bibfield  {title} {\bibinfo {title} {High-order quantum
  algorithm for solving linear differential equations},\ }\href
  {https://doi.org/10.1088/1751-8113/47/10/105301} {\bibfield  {journal}
  {\bibinfo  {journal} {Journal of Physics A: Mathematical and Theoretical}\
  }\textbf {\bibinfo {volume} {47}},\ \bibinfo {pages} {105301} (\bibinfo
  {year} {2014})}\BibitemShut {NoStop}%
\bibitem [{\citenamefont {Berry}\ \emph {et~al.}(2017)\citenamefont {Berry},
  \citenamefont {Childs}, \citenamefont {Ostrander},\ and\ \citenamefont
  {Wang}}]{berry2017}%
  \BibitemOpen
  \bibfield  {author} {\bibinfo {author} {\bibfnamefont {D.~W.}\ \bibnamefont
  {Berry}}, \bibinfo {author} {\bibfnamefont {A.~M.}\ \bibnamefont {Childs}},
  \bibinfo {author} {\bibfnamefont {A.}~\bibnamefont {Ostrander}},\ and\
  \bibinfo {author} {\bibfnamefont {G.}~\bibnamefont {Wang}},\ }\bibfield
  {title} {\bibinfo {title} {Quantum algorithm for linear differential
  equations with exponentially improved dependence on precision},\ }\href
  {https://doi.org/10.1007/s00220-017-3002-y} {\bibfield  {journal} {\bibinfo
  {journal} {Communications in Mathematical Physics}\ }\textbf {\bibinfo
  {volume} {356}},\ \bibinfo {pages} {1057} (\bibinfo {year}
  {2017})}\BibitemShut {NoStop}%
\bibitem [{\citenamefont {Xin}\ \emph {et~al.}(2020)\citenamefont {Xin},
  \citenamefont {Wei}, \citenamefont {Cui}, \citenamefont {Xiao}, \citenamefont
  {Arrazola}, \citenamefont {Lamata}, \citenamefont {Kong}, \citenamefont {Lu},
  \citenamefont {Solano},\ and\ \citenamefont {Long}}]{Solano2020}%
  \BibitemOpen
  \bibfield  {author} {\bibinfo {author} {\bibfnamefont {T.}~\bibnamefont
  {Xin}}, \bibinfo {author} {\bibfnamefont {S.}~\bibnamefont {Wei}}, \bibinfo
  {author} {\bibfnamefont {J.}~\bibnamefont {Cui}}, \bibinfo {author}
  {\bibfnamefont {J.}~\bibnamefont {Xiao}}, \bibinfo {author} {\bibfnamefont
  {I.~n.}\ \bibnamefont {Arrazola}}, \bibinfo {author} {\bibfnamefont
  {L.}~\bibnamefont {Lamata}}, \bibinfo {author} {\bibfnamefont
  {X.}~\bibnamefont {Kong}}, \bibinfo {author} {\bibfnamefont {D.}~\bibnamefont
  {Lu}}, \bibinfo {author} {\bibfnamefont {E.}~\bibnamefont {Solano}},\ and\
  \bibinfo {author} {\bibfnamefont {G.}~\bibnamefont {Long}},\ }\bibfield
  {title} {\bibinfo {title} {Quantum algorithm for solving linear differential
  equations: Theory and experiment},\ }\href
  {https://doi.org/10.1103/PhysRevA.101.032307} {\bibfield  {journal} {\bibinfo
   {journal} {Phys. Rev. A}\ }\textbf {\bibinfo {volume} {101}},\ \bibinfo
  {pages} {032307} (\bibinfo {year} {2020})}\BibitemShut {NoStop}%
\bibitem [{\citenamefont {Leyton}\ and\ \citenamefont
  {Osborne}(2008{\natexlab{a}})}]{leyton2008}%
  \BibitemOpen
  \bibfield  {author} {\bibinfo {author} {\bibfnamefont {S.~K.}\ \bibnamefont
  {Leyton}}\ and\ \bibinfo {author} {\bibfnamefont {T.~J.}\ \bibnamefont
  {Osborne}},\ }\href@noop {} {\bibinfo {title} {A quantum algorithm to solve
  nonlinear differential equations}} (\bibinfo {year} {2008}{\natexlab{a}}),\
  \Eprint {https://arxiv.org/abs/0812.4423} {arXiv:0812.4423 [quant-ph]}
  \BibitemShut {NoStop}%
\bibitem [{\citenamefont {Arrazola}\ \emph {et~al.}(2019)\citenamefont
  {Arrazola}, \citenamefont {Kalajdzievski}, \citenamefont {Weedbrook},\ and\
  \citenamefont {Lloyd}}]{Arrazola2019}%
  \BibitemOpen
  \bibfield  {author} {\bibinfo {author} {\bibfnamefont {J.~M.}\ \bibnamefont
  {Arrazola}}, \bibinfo {author} {\bibfnamefont {T.}~\bibnamefont
  {Kalajdzievski}}, \bibinfo {author} {\bibfnamefont {C.}~\bibnamefont
  {Weedbrook}},\ and\ \bibinfo {author} {\bibfnamefont {S.}~\bibnamefont
  {Lloyd}},\ }\bibfield  {title} {\bibinfo {title} {Quantum algorithm for
  nonhomogeneous linear partial differential equations},\ }\href
  {https://doi.org/10.1103/PhysRevA.100.032306} {\bibfield  {journal} {\bibinfo
   {journal} {Phys. Rev. A}\ }\textbf {\bibinfo {volume} {100}},\ \bibinfo
  {pages} {032306} (\bibinfo {year} {2019})}\BibitemShut {NoStop}%
\bibitem [{\citenamefont {Bravyi}\ \emph {et~al.}(2020)\citenamefont {Bravyi},
  \citenamefont {Gosset}, \citenamefont {Koenig},\ and\ \citenamefont
  {Tomamichel}}]{bravyi2020}%
  \BibitemOpen
  \bibfield  {author} {\bibinfo {author} {\bibfnamefont {S.}~\bibnamefont
  {Bravyi}}, \bibinfo {author} {\bibfnamefont {D.}~\bibnamefont {Gosset}},
  \bibinfo {author} {\bibfnamefont {R.}~\bibnamefont {Koenig}},\ and\ \bibinfo
  {author} {\bibfnamefont {M.}~\bibnamefont {Tomamichel}},\ }\bibfield  {title}
  {\bibinfo {title} {Quantum advantage with noisy shallow circuits},\ }\href
  {https://doi.org/https://doi.org/10.1038/s41567-020-0948-z} {\bibfield
  {journal} {\bibinfo  {journal} {Nature Physics}\ }\textbf {\bibinfo {volume}
  {16}},\ \bibinfo {pages} {1040} (\bibinfo {year} {2020})}\BibitemShut
  {NoStop}%
\bibitem [{\citenamefont {Arute}\ \emph {et~al.}(2019)\citenamefont {Arute},
  \citenamefont {Arya}, \citenamefont {Babbush}, \citenamefont {Bacon},
  \citenamefont {Bardin}, \citenamefont {Barends}, \citenamefont {Biswas},
  \citenamefont {Boixo}, \citenamefont {Brandao},\ and\ \citenamefont
  {Buell}}]{arute2019}%
  \BibitemOpen
  \bibfield  {author} {\bibinfo {author} {\bibfnamefont {F.}~\bibnamefont
  {Arute}}, \bibinfo {author} {\bibfnamefont {K.}~\bibnamefont {Arya}},
  \bibinfo {author} {\bibfnamefont {R.}~\bibnamefont {Babbush}}, \bibinfo
  {author} {\bibfnamefont {D.}~\bibnamefont {Bacon}}, \bibinfo {author}
  {\bibfnamefont {J.~C.}\ \bibnamefont {Bardin}}, \bibinfo {author}
  {\bibfnamefont {R.}~\bibnamefont {Barends}}, \bibinfo {author} {\bibfnamefont
  {R.}~\bibnamefont {Biswas}}, \bibinfo {author} {\bibfnamefont
  {S.}~\bibnamefont {Boixo}}, \bibinfo {author} {\bibfnamefont {F.~G.}\
  \bibnamefont {Brandao}},\ and\ \bibinfo {author} {\bibfnamefont {D.~A.}\
  \bibnamefont {Buell}},\ }\bibfield  {title} {\bibinfo {title} {Quantum
  supremacy using a programmable superconducting processor},\ }\href@noop {}
  {\bibfield  {journal} {\bibinfo  {journal} {Nature}\ }\textbf {\bibinfo
  {volume} {574}},\ \bibinfo {pages} {505} (\bibinfo {year}
  {2019})}\BibitemShut {NoStop}%
\bibitem [{\citenamefont {Zhong}\ \emph {et~al.}(2020)\citenamefont {Zhong},
  \citenamefont {Wang}, \citenamefont {Deng}, \citenamefont {Chen},
  \citenamefont {Peng}, \citenamefont {Luo}, \citenamefont {Qin}, \citenamefont
  {Wu}, \citenamefont {Ding},\ and\ \citenamefont {Hu}}]{zhong2020}%
  \BibitemOpen
  \bibfield  {author} {\bibinfo {author} {\bibfnamefont {H.-S.}\ \bibnamefont
  {Zhong}}, \bibinfo {author} {\bibfnamefont {H.}~\bibnamefont {Wang}},
  \bibinfo {author} {\bibfnamefont {Y.-H.}\ \bibnamefont {Deng}}, \bibinfo
  {author} {\bibfnamefont {M.-C.}\ \bibnamefont {Chen}}, \bibinfo {author}
  {\bibfnamefont {L.-C.}\ \bibnamefont {Peng}}, \bibinfo {author}
  {\bibfnamefont {Y.-H.}\ \bibnamefont {Luo}}, \bibinfo {author} {\bibfnamefont
  {J.}~\bibnamefont {Qin}}, \bibinfo {author} {\bibfnamefont {D.}~\bibnamefont
  {Wu}}, \bibinfo {author} {\bibfnamefont {X.}~\bibnamefont {Ding}},\ and\
  \bibinfo {author} {\bibfnamefont {Y.}~\bibnamefont {Hu}},\ }\bibfield
  {title} {\bibinfo {title} {Quantum computational advantage using photons},\
  }\href@noop {} {\bibfield  {journal} {\bibinfo  {journal} {Science}\ }\textbf
  {\bibinfo {volume} {370}},\ \bibinfo {pages} {1460} (\bibinfo {year}
  {2020})}\BibitemShut {NoStop}%
\bibitem [{\citenamefont {King}\ \emph {et~al.}(2021)\citenamefont {King},
  \citenamefont {Raymond}, \citenamefont {Lanting}, \citenamefont {Isakov},
  \citenamefont {Mohseni}, \citenamefont {Poulin-Lamarre}, \citenamefont
  {Ejtemaee}, \citenamefont {Bernoudy}, \citenamefont {Ozfidan},\ and\
  \citenamefont {Smirnov}}]{king2021}%
  \BibitemOpen
  \bibfield  {author} {\bibinfo {author} {\bibfnamefont {A.~D.}\ \bibnamefont
  {King}}, \bibinfo {author} {\bibfnamefont {J.}~\bibnamefont {Raymond}},
  \bibinfo {author} {\bibfnamefont {T.}~\bibnamefont {Lanting}}, \bibinfo
  {author} {\bibfnamefont {S.~V.}\ \bibnamefont {Isakov}}, \bibinfo {author}
  {\bibfnamefont {M.}~\bibnamefont {Mohseni}}, \bibinfo {author} {\bibfnamefont
  {G.}~\bibnamefont {Poulin-Lamarre}}, \bibinfo {author} {\bibfnamefont
  {S.}~\bibnamefont {Ejtemaee}}, \bibinfo {author} {\bibfnamefont
  {W.}~\bibnamefont {Bernoudy}}, \bibinfo {author} {\bibfnamefont
  {I.}~\bibnamefont {Ozfidan}},\ and\ \bibinfo {author} {\bibfnamefont {A.~Y.}\
  \bibnamefont {Smirnov}},\ }\bibfield  {title} {\bibinfo {title} {Scaling
  advantage over path-integral monte carlo in quantum simulation of
  geometrically frustrated magnets},\ }\href@noop {} {\bibfield  {journal}
  {\bibinfo  {journal} {Nature communications}\ }\textbf {\bibinfo {volume}
  {12}},\ \bibinfo {pages} {1} (\bibinfo {year} {2021})}\BibitemShut {NoStop}%
\bibitem [{\citenamefont {Verstraete}\ \emph {et~al.}(2009)\citenamefont
  {Verstraete}, \citenamefont {Wolf},\ and\ \citenamefont
  {Cirac}}]{verstraete2009}%
  \BibitemOpen
  \bibfield  {author} {\bibinfo {author} {\bibfnamefont {F.}~\bibnamefont
  {Verstraete}}, \bibinfo {author} {\bibfnamefont {M.~M.}\ \bibnamefont
  {Wolf}},\ and\ \bibinfo {author} {\bibfnamefont {J.~I.}\ \bibnamefont
  {Cirac}},\ }\bibfield  {title} {\bibinfo {title} {Quantum computation and
  quantum-state engineering driven by dissipation},\ }\href@noop {} {\bibfield
  {journal} {\bibinfo  {journal} {Nature physics}\ }\textbf {\bibinfo {volume}
  {5}},\ \bibinfo {pages} {633} (\bibinfo {year} {2009})}\BibitemShut {NoStop}%
\bibitem [{\citenamefont {Briegel}\ \emph {et~al.}(2009)\citenamefont
  {Briegel}, \citenamefont {Browne}, \citenamefont {D{\"u}r}, \citenamefont
  {Raussendorf},\ and\ \citenamefont {Van~den Nest}}]{briegel2009}%
  \BibitemOpen
  \bibfield  {author} {\bibinfo {author} {\bibfnamefont {H.~J.}\ \bibnamefont
  {Briegel}}, \bibinfo {author} {\bibfnamefont {D.~E.}\ \bibnamefont {Browne}},
  \bibinfo {author} {\bibfnamefont {W.}~\bibnamefont {D{\"u}r}}, \bibinfo
  {author} {\bibfnamefont {R.}~\bibnamefont {Raussendorf}},\ and\ \bibinfo
  {author} {\bibfnamefont {M.}~\bibnamefont {Van~den Nest}},\ }\bibfield
  {title} {\bibinfo {title} {Measurement-based quantum computation},\
  }\href@noop {} {\bibfield  {journal} {\bibinfo  {journal} {Nature Physics}\
  }\textbf {\bibinfo {volume} {5}},\ \bibinfo {pages} {19} (\bibinfo {year}
  {2009})}\BibitemShut {NoStop}%
\bibitem [{\citenamefont {Van~den Nest}\ \emph {et~al.}(2006)\citenamefont
  {Van~den Nest}, \citenamefont {Miyake}, \citenamefont {D\"ur},\ and\
  \citenamefont {Briegel}}]{van_den_nest2006}%
  \BibitemOpen
  \bibfield  {author} {\bibinfo {author} {\bibfnamefont {M.}~\bibnamefont
  {Van~den Nest}}, \bibinfo {author} {\bibfnamefont {A.}~\bibnamefont
  {Miyake}}, \bibinfo {author} {\bibfnamefont {W.}~\bibnamefont {D\"ur}},\ and\
  \bibinfo {author} {\bibfnamefont {H.~J.}\ \bibnamefont {Briegel}},\
  }\bibfield  {title} {\bibinfo {title} {Universal resources for
  measurement-based quantum computation},\ }\href
  {https://doi.org/10.1103/PhysRevLett.97.150504} {\bibfield  {journal}
  {\bibinfo  {journal} {Phys. Rev. Lett.}\ }\textbf {\bibinfo {volume} {97}},\
  \bibinfo {pages} {150504} (\bibinfo {year} {2006})}\BibitemShut {NoStop}%
\bibitem [{\citenamefont {Dong}\ \emph {et~al.}(2006)\citenamefont {Dong},
  \citenamefont {Xue}, \citenamefont {Yang},\ and\ \citenamefont
  {Cao}}]{Dong2006}%
  \BibitemOpen
  \bibfield  {author} {\bibinfo {author} {\bibfnamefont {P.}~\bibnamefont
  {Dong}}, \bibinfo {author} {\bibfnamefont {Z.-Y.}\ \bibnamefont {Xue}},
  \bibinfo {author} {\bibfnamefont {M.}~\bibnamefont {Yang}},\ and\ \bibinfo
  {author} {\bibfnamefont {Z.-L.}\ \bibnamefont {Cao}},\ }\bibfield  {title}
  {\bibinfo {title} {Generation of cluster states},\ }\href
  {https://doi.org/10.1103/PhysRevA.73.033818} {\bibfield  {journal} {\bibinfo
  {journal} {Phys. Rev. A}\ }\textbf {\bibinfo {volume} {73}},\ \bibinfo
  {pages} {033818} (\bibinfo {year} {2006})}\BibitemShut {NoStop}%
\bibitem [{\citenamefont {Raussendorf}\ \emph {et~al.}(2003)\citenamefont
  {Raussendorf}, \citenamefont {Browne},\ and\ \citenamefont
  {Briegel}}]{Raussendorf2003}%
  \BibitemOpen
  \bibfield  {author} {\bibinfo {author} {\bibfnamefont {R.}~\bibnamefont
  {Raussendorf}}, \bibinfo {author} {\bibfnamefont {D.~E.}\ \bibnamefont
  {Browne}},\ and\ \bibinfo {author} {\bibfnamefont {H.~J.}\ \bibnamefont
  {Briegel}},\ }\bibfield  {title} {\bibinfo {title} {Measurement-based quantum
  computation on cluster states},\ }\href
  {https://doi.org/10.1103/PhysRevA.68.022312} {\bibfield  {journal} {\bibinfo
  {journal} {Phys. Rev. A}\ }\textbf {\bibinfo {volume} {68}},\ \bibinfo
  {pages} {022312} (\bibinfo {year} {2003})}\BibitemShut {NoStop}%
\bibitem [{\citenamefont {Leymann}\ and\ \citenamefont
  {Barzen}(2020)}]{Leymann2020}%
  \BibitemOpen
  \bibfield  {author} {\bibinfo {author} {\bibfnamefont {F.}~\bibnamefont
  {Leymann}}\ and\ \bibinfo {author} {\bibfnamefont {J.}~\bibnamefont
  {Barzen}},\ }\bibfield  {title} {\bibinfo {title} {The bitter truth about
  gate-based quantum algorithms in the {NISQ} era},\ }\href
  {https://doi.org/10.1088/2058-9565/abae7d} {\bibfield  {journal} {\bibinfo
  {journal} {Quantum Science and Technology}\ }\textbf {\bibinfo {volume}
  {5}},\ \bibinfo {pages} {044007} (\bibinfo {year} {2020})}\BibitemShut
  {NoStop}%
\bibitem [{Note1()}]{Note1}%
  \BibitemOpen
  \bibinfo {note} {It is worth noting that for binary encoding, the operations
  on each qubit can be done in parallel for a given string, resulting in a low
  depth circuit for this case of preparation}\BibitemShut {NoStop}%
\bibitem [{\citenamefont {Jiang}\ \emph {et~al.}(2018)\citenamefont {Jiang},
  \citenamefont {Britt}, \citenamefont {McCaskey}, \citenamefont {Humble},\
  and\ \citenamefont {Kais}}]{jiang2018}%
  \BibitemOpen
  \bibfield  {author} {\bibinfo {author} {\bibfnamefont {S.}~\bibnamefont
  {Jiang}}, \bibinfo {author} {\bibfnamefont {K.~A.}\ \bibnamefont {Britt}},
  \bibinfo {author} {\bibfnamefont {A.~J.}\ \bibnamefont {McCaskey}}, \bibinfo
  {author} {\bibfnamefont {T.~S.}\ \bibnamefont {Humble}},\ and\ \bibinfo
  {author} {\bibfnamefont {S.}~\bibnamefont {Kais}},\ }\bibfield  {title}
  {\bibinfo {title} {Quantum annealing for prime factorization},\ }\href@noop
  {} {\bibfield  {journal} {\bibinfo  {journal} {Scientific reports}\ }\textbf
  {\bibinfo {volume} {8}},\ \bibinfo {pages} {1} (\bibinfo {year}
  {2018})}\BibitemShut {NoStop}%
\bibitem [{\citenamefont {Teplukhin}\ \emph {et~al.}(2020)\citenamefont
  {Teplukhin}, \citenamefont {Kendrick}, \citenamefont {Tretiak},\ and\
  \citenamefont {Dub}}]{teplukhin2020}%
  \BibitemOpen
  \bibfield  {author} {\bibinfo {author} {\bibfnamefont {A.}~\bibnamefont
  {Teplukhin}}, \bibinfo {author} {\bibfnamefont {B.~K.}\ \bibnamefont
  {Kendrick}}, \bibinfo {author} {\bibfnamefont {S.}~\bibnamefont {Tretiak}},\
  and\ \bibinfo {author} {\bibfnamefont {P.~A.}\ \bibnamefont {Dub}},\
  }\bibfield  {title} {\bibinfo {title} {Electronic structure with direct
  diagonalization on a d-wave quantum annealer},\ }\href@noop {} {\bibfield
  {journal} {\bibinfo  {journal} {Scientific reports}\ }\textbf {\bibinfo
  {volume} {10}},\ \bibinfo {pages} {1} (\bibinfo {year} {2020})}\BibitemShut
  {NoStop}%
\bibitem [{\citenamefont {Shende}\ and\ \citenamefont
  {Markov}(2004)}]{shende2004}%
  \BibitemOpen
  \bibfield  {author} {\bibinfo {author} {\bibfnamefont {V.~V.}\ \bibnamefont
  {Shende}}\ and\ \bibinfo {author} {\bibfnamefont {I.~L.}\ \bibnamefont
  {Markov}},\ }\bibfield  {title} {\bibinfo {title} {Quantum circuits for
  incompletely specified two-qubit operators},\ }\href@noop {} {\bibfield
  {journal} {\bibinfo  {journal} {arXiv preprint quant-ph/0401162}\ } (\bibinfo
  {year} {2004})}\BibitemShut {NoStop}%
\bibitem [{\citenamefont {Long}\ and\ \citenamefont {Sun}(2001)}]{long2001}%
  \BibitemOpen
  \bibfield  {author} {\bibinfo {author} {\bibfnamefont {G.-L.}\ \bibnamefont
  {Long}}\ and\ \bibinfo {author} {\bibfnamefont {Y.}~\bibnamefont {Sun}},\
  }\bibfield  {title} {\bibinfo {title} {Efficient scheme for initializing a
  quantum register with an arbitrary superposed state},\ }\href
  {https://doi.org/10.1103/PhysRevA.64.014303} {\bibfield  {journal} {\bibinfo
  {journal} {Phys. Rev. A}\ }\textbf {\bibinfo {volume} {64}},\ \bibinfo
  {pages} {014303} (\bibinfo {year} {2001})}\BibitemShut {NoStop}%
\bibitem [{\citenamefont {Barenco}\ \emph {et~al.}(1995)\citenamefont
  {Barenco}, \citenamefont {Bennett}, \citenamefont {Cleve}, \citenamefont
  {DiVincenzo}, \citenamefont {Margolus}, \citenamefont {Shor}, \citenamefont
  {Sleator}, \citenamefont {Smolin},\ and\ \citenamefont
  {Weinfurter}}]{Barenco1995}%
  \BibitemOpen
  \bibfield  {author} {\bibinfo {author} {\bibfnamefont {A.}~\bibnamefont
  {Barenco}}, \bibinfo {author} {\bibfnamefont {C.~H.}\ \bibnamefont
  {Bennett}}, \bibinfo {author} {\bibfnamefont {R.}~\bibnamefont {Cleve}},
  \bibinfo {author} {\bibfnamefont {D.~P.}\ \bibnamefont {DiVincenzo}},
  \bibinfo {author} {\bibfnamefont {N.}~\bibnamefont {Margolus}}, \bibinfo
  {author} {\bibfnamefont {P.}~\bibnamefont {Shor}}, \bibinfo {author}
  {\bibfnamefont {T.}~\bibnamefont {Sleator}}, \bibinfo {author} {\bibfnamefont
  {J.~A.}\ \bibnamefont {Smolin}},\ and\ \bibinfo {author} {\bibfnamefont
  {H.}~\bibnamefont {Weinfurter}},\ }\bibfield  {title} {\bibinfo {title}
  {Elementary gates for quantum computation},\ }\href
  {https://doi.org/10.1103/PhysRevA.52.3457} {\bibfield  {journal} {\bibinfo
  {journal} {Phys. Rev. A}\ }\textbf {\bibinfo {volume} {52}},\ \bibinfo
  {pages} {3457} (\bibinfo {year} {1995})}\BibitemShut {NoStop}%
\bibitem [{\citenamefont {Soklakov}\ and\ \citenamefont
  {Schack}(2006)}]{Soklakov2006}%
  \BibitemOpen
  \bibfield  {author} {\bibinfo {author} {\bibfnamefont {A.~N.}\ \bibnamefont
  {Soklakov}}\ and\ \bibinfo {author} {\bibfnamefont {R.}~\bibnamefont
  {Schack}},\ }\bibfield  {title} {\bibinfo {title} {Efficient state
  preparation for a register of quantum bits},\ }\href
  {https://doi.org/10.1103/PhysRevA.73.012307} {\bibfield  {journal} {\bibinfo
  {journal} {Phys. Rev. A}\ }\textbf {\bibinfo {volume} {73}},\ \bibinfo
  {pages} {012307} (\bibinfo {year} {2006})}\BibitemShut {NoStop}%
\bibitem [{\citenamefont {Giovannetti}\ \emph
  {et~al.}(2008{\natexlab{a}})\citenamefont {Giovannetti}, \citenamefont
  {Lloyd},\ and\ \citenamefont {Maccone}}]{giovannetti2008}%
  \BibitemOpen
  \bibfield  {author} {\bibinfo {author} {\bibfnamefont {V.}~\bibnamefont
  {Giovannetti}}, \bibinfo {author} {\bibfnamefont {S.}~\bibnamefont {Lloyd}},\
  and\ \bibinfo {author} {\bibfnamefont {L.}~\bibnamefont {Maccone}},\
  }\bibfield  {title} {\bibinfo {title} {Quantum random access memory},\ }\href
  {https://doi.org/10.1103/PhysRevLett.100.160501} {\bibfield  {journal}
  {\bibinfo  {journal} {Phys. Rev. Lett.}\ }\textbf {\bibinfo {volume} {100}},\
  \bibinfo {pages} {160501} (\bibinfo {year} {2008}{\natexlab{a}})}\BibitemShut
  {NoStop}%
\bibitem [{\citenamefont {Ventura}\ and\ \citenamefont
  {Martinez}(2000)}]{ventura2000}%
  \BibitemOpen
  \bibfield  {author} {\bibinfo {author} {\bibfnamefont {D.}~\bibnamefont
  {Ventura}}\ and\ \bibinfo {author} {\bibfnamefont {T.}~\bibnamefont
  {Martinez}},\ }\bibfield  {title} {\bibinfo {title} {Quantum associative
  memory},\ }\href
  {https://doi.org/https://doi.org/10.1016/S0020-0255(99)00101-2} {\bibfield
  {journal} {\bibinfo  {journal} {Information Sciences}\ }\textbf {\bibinfo
  {volume} {124}},\ \bibinfo {pages} {273} (\bibinfo {year}
  {2000})}\BibitemShut {NoStop}%
\bibitem [{\citenamefont {Giovannetti}\ \emph
  {et~al.}(2008{\natexlab{b}})\citenamefont {Giovannetti}, \citenamefont
  {Lloyd},\ and\ \citenamefont {Maccone}}]{giovannetti2008arch}%
  \BibitemOpen
  \bibfield  {author} {\bibinfo {author} {\bibfnamefont {V.}~\bibnamefont
  {Giovannetti}}, \bibinfo {author} {\bibfnamefont {S.}~\bibnamefont {Lloyd}},\
  and\ \bibinfo {author} {\bibfnamefont {L.}~\bibnamefont {Maccone}},\
  }\bibfield  {title} {\bibinfo {title} {Architectures for a quantum random
  access memory},\ }\href {https://doi.org/10.1103/PhysRevA.78.052310}
  {\bibfield  {journal} {\bibinfo  {journal} {Phys. Rev. A}\ }\textbf {\bibinfo
  {volume} {78}},\ \bibinfo {pages} {052310} (\bibinfo {year}
  {2008}{\natexlab{b}})}\BibitemShut {NoStop}%
\bibitem [{\citenamefont {Park}\ \emph {et~al.}(2019)\citenamefont {Park},
  \citenamefont {Petruccione},\ and\ \citenamefont {Rhee}}]{park2019}%
  \BibitemOpen
  \bibfield  {author} {\bibinfo {author} {\bibfnamefont {D.}~\bibnamefont
  {Park}}, \bibinfo {author} {\bibfnamefont {F.}~\bibnamefont {Petruccione}},\
  and\ \bibinfo {author} {\bibfnamefont {J.}~\bibnamefont {Rhee}},\ }\bibfield
  {title} {\bibinfo {title} {Circuit-based quantum random access memory for
  classical data},\ }\bibfield  {journal} {\bibinfo  {journal} {Sci Rep}\
  }\textbf {\bibinfo {volume} {3949}},\ \href
  {https://doi.org/10.1038/s41598-019-40439-3} {10.1038/s41598-019-40439-3}
  (\bibinfo {year} {2019})\BibitemShut {NoStop}%
\bibitem [{\citenamefont {Paetznick}\ and\ \citenamefont
  {Reichardt}(2013)}]{Paetznick2013}%
  \BibitemOpen
  \bibfield  {author} {\bibinfo {author} {\bibfnamefont {A.}~\bibnamefont
  {Paetznick}}\ and\ \bibinfo {author} {\bibfnamefont {B.~W.}\ \bibnamefont
  {Reichardt}},\ }\bibfield  {title} {\bibinfo {title} {Universal
  fault-tolerant quantum computation with only transversal gates and error
  correction},\ }\href {https://doi.org/10.1103/PhysRevLett.111.090505}
  {\bibfield  {journal} {\bibinfo  {journal} {Phys. Rev. Lett.}\ }\textbf
  {\bibinfo {volume} {111}},\ \bibinfo {pages} {090505} (\bibinfo {year}
  {2013})}\BibitemShut {NoStop}%
\bibitem [{\citenamefont {Anderson}\ \emph {et~al.}(2014)\citenamefont
  {Anderson}, \citenamefont {Duclos-Cianci},\ and\ \citenamefont
  {Poulin}}]{anderson2014}%
  \BibitemOpen
  \bibfield  {author} {\bibinfo {author} {\bibfnamefont {J.~T.}\ \bibnamefont
  {Anderson}}, \bibinfo {author} {\bibfnamefont {G.}~\bibnamefont
  {Duclos-Cianci}},\ and\ \bibinfo {author} {\bibfnamefont {D.}~\bibnamefont
  {Poulin}},\ }\bibfield  {title} {\bibinfo {title} {Fault-tolerant conversion
  between the steane and reed-muller quantum codes},\ }\href
  {https://doi.org/10.1103/PhysRevLett.113.080501} {\bibfield  {journal}
  {\bibinfo  {journal} {Phys. Rev. Lett.}\ }\textbf {\bibinfo {volume} {113}},\
  \bibinfo {pages} {080501} (\bibinfo {year} {2014})}\BibitemShut {NoStop}%
\bibitem [{\citenamefont {Jochym-O'Connor}\ and\ \citenamefont
  {Laflamme}(2014)}]{Oconnor2014}%
  \BibitemOpen
  \bibfield  {author} {\bibinfo {author} {\bibfnamefont {T.}~\bibnamefont
  {Jochym-O'Connor}}\ and\ \bibinfo {author} {\bibfnamefont {R.}~\bibnamefont
  {Laflamme}},\ }\bibfield  {title} {\bibinfo {title} {Using concatenated
  quantum codes for universal fault-tolerant quantum gates},\ }\href
  {https://doi.org/10.1103/PhysRevLett.112.010505} {\bibfield  {journal}
  {\bibinfo  {journal} {Phys. Rev. Lett.}\ }\textbf {\bibinfo {volume} {112}},\
  \bibinfo {pages} {010505} (\bibinfo {year} {2014})}\BibitemShut {NoStop}%
\bibitem [{\citenamefont {Mohseni}\ \emph {et~al.}(2008)\citenamefont
  {Mohseni}, \citenamefont {Rezakhani},\ and\ \citenamefont
  {Lidar}}]{mohseni2008}%
  \BibitemOpen
  \bibfield  {author} {\bibinfo {author} {\bibfnamefont {M.}~\bibnamefont
  {Mohseni}}, \bibinfo {author} {\bibfnamefont {A.~T.}\ \bibnamefont
  {Rezakhani}},\ and\ \bibinfo {author} {\bibfnamefont {D.~A.}\ \bibnamefont
  {Lidar}},\ }\bibfield  {title} {\bibinfo {title} {Quantum-process tomography:
  Resource analysis of different strategies},\ }\href
  {https://doi.org/10.1103/PhysRevA.77.032322} {\bibfield  {journal} {\bibinfo
  {journal} {Phys. Rev. A}\ }\textbf {\bibinfo {volume} {77}},\ \bibinfo
  {pages} {032322} (\bibinfo {year} {2008})}\BibitemShut {NoStop}%
\bibitem [{\citenamefont {Aaronson}(2015)}]{aaronson2015read}%
  \BibitemOpen
  \bibfield  {author} {\bibinfo {author} {\bibfnamefont {S.}~\bibnamefont
  {Aaronson}},\ }\bibfield  {title} {\bibinfo {title} {Read the fine print},\
  }\href@noop {} {\bibfield  {journal} {\bibinfo  {journal} {Nature Physics}\
  }\textbf {\bibinfo {volume} {11}},\ \bibinfo {pages} {291} (\bibinfo {year}
  {2015})}\BibitemShut {NoStop}%
\bibitem [{\citenamefont {Ekert}\ \emph {et~al.}(2002)\citenamefont {Ekert},
  \citenamefont {Alves}, \citenamefont {Oi}, \citenamefont {Horodecki},
  \citenamefont {Horodecki},\ and\ \citenamefont {Kwek}}]{Ekert2002}%
  \BibitemOpen
  \bibfield  {author} {\bibinfo {author} {\bibfnamefont {A.~K.}\ \bibnamefont
  {Ekert}}, \bibinfo {author} {\bibfnamefont {C.~M.}\ \bibnamefont {Alves}},
  \bibinfo {author} {\bibfnamefont {D.~K.~L.}\ \bibnamefont {Oi}}, \bibinfo
  {author} {\bibfnamefont {M.}~\bibnamefont {Horodecki}}, \bibinfo {author}
  {\bibfnamefont {P.}~\bibnamefont {Horodecki}},\ and\ \bibinfo {author}
  {\bibfnamefont {L.~C.}\ \bibnamefont {Kwek}},\ }\bibfield  {title} {\bibinfo
  {title} {Direct estimations of linear and nonlinear functionals of a quantum
  state},\ }\href {https://doi.org/10.1103/PhysRevLett.88.217901} {\bibfield
  {journal} {\bibinfo  {journal} {Phys. Rev. Lett.}\ }\textbf {\bibinfo
  {volume} {88}},\ \bibinfo {pages} {217901} (\bibinfo {year}
  {2002})}\BibitemShut {NoStop}%
\bibitem [{\citenamefont {Clader}\ \emph {et~al.}(2013)\citenamefont {Clader},
  \citenamefont {Jacobs},\ and\ \citenamefont {Sprouse}}]{Clader2013}%
  \BibitemOpen
  \bibfield  {author} {\bibinfo {author} {\bibfnamefont {B.~D.}\ \bibnamefont
  {Clader}}, \bibinfo {author} {\bibfnamefont {B.~C.}\ \bibnamefont {Jacobs}},\
  and\ \bibinfo {author} {\bibfnamefont {C.~R.}\ \bibnamefont {Sprouse}},\
  }\bibfield  {title} {\bibinfo {title} {Preconditioned quantum linear system
  algorithm},\ }\href {https://doi.org/10.1103/PhysRevLett.110.250504}
  {\bibfield  {journal} {\bibinfo  {journal} {Phys. Rev. Lett.}\ }\textbf
  {\bibinfo {volume} {110}},\ \bibinfo {pages} {250504} (\bibinfo {year}
  {2013})}\BibitemShut {NoStop}%
\bibitem [{\citenamefont {Cao}\ \emph {et~al.}(2013)\citenamefont {Cao},
  \citenamefont {Papageorgiou}, \citenamefont {Petras}, \citenamefont {Traub},\
  and\ \citenamefont {Kais}}]{cao2013quantum}%
  \BibitemOpen
  \bibfield  {author} {\bibinfo {author} {\bibfnamefont {Y.}~\bibnamefont
  {Cao}}, \bibinfo {author} {\bibfnamefont {A.}~\bibnamefont {Papageorgiou}},
  \bibinfo {author} {\bibfnamefont {I.}~\bibnamefont {Petras}}, \bibinfo
  {author} {\bibfnamefont {J.}~\bibnamefont {Traub}},\ and\ \bibinfo {author}
  {\bibfnamefont {S.}~\bibnamefont {Kais}},\ }\bibfield  {title} {\bibinfo
  {title} {Quantum algorithm and circuit design solving the poisson equation},\
  }\href@noop {} {\bibfield  {journal} {\bibinfo  {journal} {New Journal of
  Physics}\ }\textbf {\bibinfo {volume} {15}},\ \bibinfo {pages} {013021}
  (\bibinfo {year} {2013})}\BibitemShut {NoStop}%
\bibitem [{\citenamefont {Montanaro}\ and\ \citenamefont
  {Pallister}(2016)}]{montanaro2016quantum}%
  \BibitemOpen
  \bibfield  {author} {\bibinfo {author} {\bibfnamefont {A.}~\bibnamefont
  {Montanaro}}\ and\ \bibinfo {author} {\bibfnamefont {S.}~\bibnamefont
  {Pallister}},\ }\bibfield  {title} {\bibinfo {title} {Quantum algorithms and
  the finite element method},\ }\href@noop {} {\bibfield  {journal} {\bibinfo
  {journal} {Physical Review A}\ }\textbf {\bibinfo {volume} {93}},\ \bibinfo
  {pages} {032324} (\bibinfo {year} {2016})}\BibitemShut {NoStop}%
\bibitem [{\citenamefont {Costa}\ \emph {et~al.}(2019)\citenamefont {Costa},
  \citenamefont {Jordan},\ and\ \citenamefont {Ostrander}}]{costa2019quantum}%
  \BibitemOpen
  \bibfield  {author} {\bibinfo {author} {\bibfnamefont {P.~C.}\ \bibnamefont
  {Costa}}, \bibinfo {author} {\bibfnamefont {S.}~\bibnamefont {Jordan}},\ and\
  \bibinfo {author} {\bibfnamefont {A.}~\bibnamefont {Ostrander}},\ }\bibfield
  {title} {\bibinfo {title} {Quantum algorithm for simulating the wave
  equation},\ }\href@noop {} {\bibfield  {journal} {\bibinfo  {journal}
  {Physical Review A}\ }\textbf {\bibinfo {volume} {99}},\ \bibinfo {pages}
  {012323} (\bibinfo {year} {2019})}\BibitemShut {NoStop}%
\bibitem [{\citenamefont {Fillion-Gourdeau}\ and\ \citenamefont
  {Lorin}(2019)}]{Fillion2019}%
  \BibitemOpen
  \bibfield  {author} {\bibinfo {author} {\bibfnamefont {F.}~\bibnamefont
  {Fillion-Gourdeau}}\ and\ \bibinfo {author} {\bibfnamefont {E.}~\bibnamefont
  {Lorin}},\ }\bibfield  {title} {\bibinfo {title} {Simple digital quantum
  algorithm for symmetric first-order linear hyperbolic systems},\ }\bibfield
  {journal} {\bibinfo  {journal} {Numer. Algor.}\ }\textbf {\bibinfo {volume}
  {82}},\ \href {https://doi.org/https://doi.org/10.1007/s11075-018-0639-3}
  {https://doi.org/10.1007/s11075-018-0639-3} (\bibinfo {year}
  {2019})\BibitemShut {NoStop}%
\bibitem [{\citenamefont {Wang}\ \emph {et~al.}(2020)\citenamefont {Wang},
  \citenamefont {Wang}, \citenamefont {Li}, \citenamefont {Fan}, \citenamefont
  {Wei},\ and\ \citenamefont {Gu}}]{wang2020quantum}%
  \BibitemOpen
  \bibfield  {author} {\bibinfo {author} {\bibfnamefont {S.}~\bibnamefont
  {Wang}}, \bibinfo {author} {\bibfnamefont {Z.}~\bibnamefont {Wang}}, \bibinfo
  {author} {\bibfnamefont {W.}~\bibnamefont {Li}}, \bibinfo {author}
  {\bibfnamefont {L.}~\bibnamefont {Fan}}, \bibinfo {author} {\bibfnamefont
  {Z.}~\bibnamefont {Wei}},\ and\ \bibinfo {author} {\bibfnamefont
  {Y.}~\bibnamefont {Gu}},\ }\bibfield  {title} {\bibinfo {title} {Quantum fast
  poisson solver: the algorithm and complete and modular circuit design},\
  }\href@noop {} {\bibfield  {journal} {\bibinfo  {journal} {Quantum
  Information Processing}\ }\textbf {\bibinfo {volume} {19}},\ \bibinfo {pages}
  {1} (\bibinfo {year} {2020})}\BibitemShut {NoStop}%
\bibitem [{\citenamefont {Leyton}\ and\ \citenamefont
  {Osborne}(2008{\natexlab{b}})}]{leyton2008quantum}%
  \BibitemOpen
  \bibfield  {author} {\bibinfo {author} {\bibfnamefont {S.~K.}\ \bibnamefont
  {Leyton}}\ and\ \bibinfo {author} {\bibfnamefont {T.~J.}\ \bibnamefont
  {Osborne}},\ }\bibfield  {title} {\bibinfo {title} {A quantum algorithm to
  solve nonlinear differential equations},\ }\href@noop {} {\bibfield
  {journal} {\bibinfo  {journal} {arXiv preprint arXiv:0812.4423}\ } (\bibinfo
  {year} {2008}{\natexlab{b}})}\BibitemShut {NoStop}%
\bibitem [{\citenamefont {Childs}\ \emph {et~al.}(2017)\citenamefont {Childs},
  \citenamefont {Kothari},\ and\ \citenamefont {Somma}}]{childs2017quantum}%
  \BibitemOpen
  \bibfield  {author} {\bibinfo {author} {\bibfnamefont {A.~M.}\ \bibnamefont
  {Childs}}, \bibinfo {author} {\bibfnamefont {R.}~\bibnamefont {Kothari}},\
  and\ \bibinfo {author} {\bibfnamefont {R.~D.}\ \bibnamefont {Somma}},\
  }\bibfield  {title} {\bibinfo {title} {Quantum algorithm for systems of
  linear equations with exponentially improved dependence on precision},\
  }\href@noop {} {\bibfield  {journal} {\bibinfo  {journal} {SIAM Journal on
  Computing}\ }\textbf {\bibinfo {volume} {46}},\ \bibinfo {pages} {1920}
  (\bibinfo {year} {2017})}\BibitemShut {NoStop}%
\bibitem [{\citenamefont {Suba{\c{s}}{\i}}\ \emph {et~al.}(2019)\citenamefont
  {Suba{\c{s}}{\i}}, \citenamefont {Somma},\ and\ \citenamefont
  {Orsucci}}]{subacsi2019quantum}%
  \BibitemOpen
  \bibfield  {author} {\bibinfo {author} {\bibfnamefont {Y.}~\bibnamefont
  {Suba{\c{s}}{\i}}}, \bibinfo {author} {\bibfnamefont {R.~D.}\ \bibnamefont
  {Somma}},\ and\ \bibinfo {author} {\bibfnamefont {D.}~\bibnamefont
  {Orsucci}},\ }\bibfield  {title} {\bibinfo {title} {Quantum algorithms for
  systems of linear equations inspired by adiabatic quantum computing},\
  }\href@noop {} {\bibfield  {journal} {\bibinfo  {journal} {Physical review
  letters}\ }\textbf {\bibinfo {volume} {122}},\ \bibinfo {pages} {060504}
  (\bibinfo {year} {2019})}\BibitemShut {NoStop}%
\bibitem [{\citenamefont {Linden}\ \emph {et~al.}(2020)\citenamefont {Linden},
  \citenamefont {Montanaro},\ and\ \citenamefont {Shao}}]{linden2020quantum}%
  \BibitemOpen
  \bibfield  {author} {\bibinfo {author} {\bibfnamefont {N.}~\bibnamefont
  {Linden}}, \bibinfo {author} {\bibfnamefont {A.}~\bibnamefont {Montanaro}},\
  and\ \bibinfo {author} {\bibfnamefont {C.}~\bibnamefont {Shao}},\ }\bibfield
  {title} {\bibinfo {title} {Quantum vs. classical algorithms for solving the
  heat equation},\ }\href@noop {} {\bibfield  {journal} {\bibinfo  {journal}
  {arXiv preprint arXiv:2004.06516}\ } (\bibinfo {year} {2020})}\BibitemShut
  {NoStop}%
\bibitem [{Note2()}]{Note2}%
  \BibitemOpen
  \bibinfo {note} {Although reference \cite {Altepeter2003} discusses quantum
  process tomography, a QST procedure is needed in order to complete the
  protocol in SQPT and AAPT schemes, and an insight about the complexity of
  quantum state tomography can be obtained.}\BibitemShut {Stop}%
\bibitem [{\citenamefont {Altepeter}\ \emph {et~al.}(2003)\citenamefont
  {Altepeter}, \citenamefont {Branning}, \citenamefont {Jeffrey}, \citenamefont
  {Wei}, \citenamefont {Kwiat}, \citenamefont {Thew}, \citenamefont {O'Brien},
  \citenamefont {Nielsen},\ and\ \citenamefont {White}}]{Altepeter2003}%
  \BibitemOpen
  \bibfield  {author} {\bibinfo {author} {\bibfnamefont {J.~B.}\ \bibnamefont
  {Altepeter}}, \bibinfo {author} {\bibfnamefont {D.}~\bibnamefont {Branning}},
  \bibinfo {author} {\bibfnamefont {E.}~\bibnamefont {Jeffrey}}, \bibinfo
  {author} {\bibfnamefont {T.~C.}\ \bibnamefont {Wei}}, \bibinfo {author}
  {\bibfnamefont {P.~G.}\ \bibnamefont {Kwiat}}, \bibinfo {author}
  {\bibfnamefont {R.~T.}\ \bibnamefont {Thew}}, \bibinfo {author}
  {\bibfnamefont {J.~L.}\ \bibnamefont {O'Brien}}, \bibinfo {author}
  {\bibfnamefont {M.~A.}\ \bibnamefont {Nielsen}},\ and\ \bibinfo {author}
  {\bibfnamefont {A.~G.}\ \bibnamefont {White}},\ }\bibfield  {title} {\bibinfo
  {title} {Ancilla-assisted quantum process tomography},\ }\href
  {https://doi.org/10.1103/PhysRevLett.90.193601} {\bibfield  {journal}
  {\bibinfo  {journal} {Phys. Rev. Lett.}\ }\textbf {\bibinfo {volume} {90}},\
  \bibinfo {pages} {193601} (\bibinfo {year} {2003})}\BibitemShut {NoStop}%
\bibitem [{\citenamefont {Qi}\ \emph {et~al.}(2013)\citenamefont {Qi},
  \citenamefont {Hou}, \citenamefont {Li}, \citenamefont {Dong}, \citenamefont
  {Xiang},\ and\ \citenamefont {Guo}}]{qi2013}%
  \BibitemOpen
  \bibfield  {author} {\bibinfo {author} {\bibfnamefont {B.}~\bibnamefont
  {Qi}}, \bibinfo {author} {\bibfnamefont {Z.}~\bibnamefont {Hou}}, \bibinfo
  {author} {\bibfnamefont {L.}~\bibnamefont {Li}}, \bibinfo {author}
  {\bibfnamefont {D.}~\bibnamefont {Dong}}, \bibinfo {author} {\bibfnamefont
  {G.}~\bibnamefont {Xiang}},\ and\ \bibinfo {author} {\bibfnamefont
  {G.}~\bibnamefont {Guo}},\ }\bibfield  {title} {\bibinfo {title} {Quantum
  state tomography via linear regression estimation},\ }\href@noop {}
  {\bibfield  {journal} {\bibinfo  {journal} {Scientific reports}\ }\textbf
  {\bibinfo {volume} {3}},\ \bibinfo {pages} {1} (\bibinfo {year}
  {2013})}\BibitemShut {NoStop}%
\bibitem [{\citenamefont {Gross}\ \emph {et~al.}(2010)\citenamefont {Gross},
  \citenamefont {Liu}, \citenamefont {Flammia}, \citenamefont {Becker},\ and\
  \citenamefont {Eisert}}]{Gross2010}%
  \BibitemOpen
  \bibfield  {author} {\bibinfo {author} {\bibfnamefont {D.}~\bibnamefont
  {Gross}}, \bibinfo {author} {\bibfnamefont {Y.-K.}\ \bibnamefont {Liu}},
  \bibinfo {author} {\bibfnamefont {S.~T.}\ \bibnamefont {Flammia}}, \bibinfo
  {author} {\bibfnamefont {S.}~\bibnamefont {Becker}},\ and\ \bibinfo {author}
  {\bibfnamefont {J.}~\bibnamefont {Eisert}},\ }\bibfield  {title} {\bibinfo
  {title} {Quantum state tomography via compressed sensing},\ }\href
  {https://doi.org/10.1103/PhysRevLett.105.150401} {\bibfield  {journal}
  {\bibinfo  {journal} {Phys. Rev. Lett.}\ }\textbf {\bibinfo {volume} {105}},\
  \bibinfo {pages} {150401} (\bibinfo {year} {2010})}\BibitemShut {NoStop}%
\bibitem [{\citenamefont {Lloyd}\ \emph {et~al.}(2014)\citenamefont {Lloyd},
  \citenamefont {Mohseni},\ and\ \citenamefont {Rebentrost}}]{lloyd2014}%
  \BibitemOpen
  \bibfield  {author} {\bibinfo {author} {\bibfnamefont {S.}~\bibnamefont
  {Lloyd}}, \bibinfo {author} {\bibfnamefont {M.}~\bibnamefont {Mohseni}},\
  and\ \bibinfo {author} {\bibfnamefont {P.}~\bibnamefont {Rebentrost}},\
  }\bibfield  {title} {\bibinfo {title} {Quantum principal component
  analysis},\ }\href {https://doi.org/10.1038/nphys3029} {\bibfield  {journal}
  {\bibinfo  {journal} {Nature Physics}\ }\textbf {\bibinfo {volume} {10}},\
  \bibinfo {pages} {631} (\bibinfo {year} {2014})}\BibitemShut {NoStop}%
\bibitem [{\citenamefont {Cramer}\ \emph {et~al.}(2010)\citenamefont {Cramer},
  \citenamefont {Plenio}, \citenamefont {Flammia}, \citenamefont {Somma},
  \citenamefont {Gross}, \citenamefont {Bartlett}, \citenamefont
  {Landon-Cardinal}, \citenamefont {Poulin},\ and\ \citenamefont
  {Liu}}]{cramer2010efficient}%
  \BibitemOpen
  \bibfield  {author} {\bibinfo {author} {\bibfnamefont {M.}~\bibnamefont
  {Cramer}}, \bibinfo {author} {\bibfnamefont {M.~B.}\ \bibnamefont {Plenio}},
  \bibinfo {author} {\bibfnamefont {S.~T.}\ \bibnamefont {Flammia}}, \bibinfo
  {author} {\bibfnamefont {R.}~\bibnamefont {Somma}}, \bibinfo {author}
  {\bibfnamefont {D.}~\bibnamefont {Gross}}, \bibinfo {author} {\bibfnamefont
  {S.~D.}\ \bibnamefont {Bartlett}}, \bibinfo {author} {\bibfnamefont
  {O.}~\bibnamefont {Landon-Cardinal}}, \bibinfo {author} {\bibfnamefont
  {D.}~\bibnamefont {Poulin}},\ and\ \bibinfo {author} {\bibfnamefont {Y.-K.}\
  \bibnamefont {Liu}},\ }\bibfield  {title} {\bibinfo {title} {Efficient
  quantum state tomography},\ }\href@noop {} {\bibfield  {journal} {\bibinfo
  {journal} {Nature communications}\ }\textbf {\bibinfo {volume} {1}},\
  \bibinfo {pages} {1} (\bibinfo {year} {2010})}\BibitemShut {NoStop}%
\bibitem [{\citenamefont {Flammia}\ \emph {et~al.}(2012)\citenamefont
  {Flammia}, \citenamefont {Gross}, \citenamefont {Liu},\ and\ \citenamefont
  {Eisert}}]{flammia2012}%
  \BibitemOpen
  \bibfield  {author} {\bibinfo {author} {\bibfnamefont {S.~T.}\ \bibnamefont
  {Flammia}}, \bibinfo {author} {\bibfnamefont {D.}~\bibnamefont {Gross}},
  \bibinfo {author} {\bibfnamefont {Y.-K.}\ \bibnamefont {Liu}},\ and\ \bibinfo
  {author} {\bibfnamefont {J.}~\bibnamefont {Eisert}},\ }\bibfield  {title}
  {\bibinfo {title} {Quantum tomography via compressed sensing: error bounds,
  sample complexity and efficient estimators},\ }\href@noop {} {\bibfield
  {journal} {\bibinfo  {journal} {New Journal of Physics}\ }\textbf {\bibinfo
  {volume} {14}},\ \bibinfo {pages} {095022} (\bibinfo {year}
  {2012})}\BibitemShut {NoStop}%
\bibitem [{\citenamefont {Aaronson}(2007)}]{aaronson2007}%
  \BibitemOpen
  \bibfield  {author} {\bibinfo {author} {\bibfnamefont {S.}~\bibnamefont
  {Aaronson}},\ }\bibfield  {title} {\bibinfo {title} {The learnability of
  quantum states},\ }\href {https://doi.org/10.1098/rspa.2007.0113} {\bibfield
  {journal} {\bibinfo  {journal} {Proceedings of the Royal Society A:
  Mathematical, Physical and Engineering Sciences}\ }\textbf {\bibinfo {volume}
  {463}},\ \bibinfo {pages} {3089} (\bibinfo {year} {2007})}\BibitemShut
  {NoStop}%
\bibitem [{Note3()}]{Note3}%
  \BibitemOpen
  \bibinfo {note} {The overall complexity is defined as in \cite {mohseni2008},
  given by the number of copies of $\rho $ times the number of gates per
  measurement.}\BibitemShut {Stop}%
\bibitem [{\citenamefont {Suzuki}\ \emph {et~al.}(2020)\citenamefont {Suzuki},
  \citenamefont {Uno}, \citenamefont {Raymond}, \citenamefont {Tanaka},
  \citenamefont {Onodera},\ and\ \citenamefont {Yamamoto}}]{Suzuki2020}%
  \BibitemOpen
  \bibfield  {author} {\bibinfo {author} {\bibfnamefont {Y.}~\bibnamefont
  {Suzuki}}, \bibinfo {author} {\bibfnamefont {S.}~\bibnamefont {Uno}},
  \bibinfo {author} {\bibfnamefont {R.}~\bibnamefont {Raymond}}, \bibinfo
  {author} {\bibfnamefont {T.}~\bibnamefont {Tanaka}}, \bibinfo {author}
  {\bibfnamefont {T.}~\bibnamefont {Onodera}},\ and\ \bibinfo {author}
  {\bibfnamefont {N.}~\bibnamefont {Yamamoto}},\ }\bibfield  {title} {\bibinfo
  {title} {Amplitude estimation without phase estimation},\ }\href@noop {}
  {\bibfield  {journal} {\bibinfo  {journal} {Quantum Information Processing}\
  }\textbf {\bibinfo {volume} {19}} (\bibinfo {year} {2020})}\BibitemShut
  {NoStop}%
\bibitem [{Note4()}]{Note4}%
  \BibitemOpen
  \bibinfo {note} {This exactly means that $|\protect \overline
  {p}_m-p_m|/|p_m|\leq \Delta $ with probability $1-\epsilon $.}\BibitemShut
  {Stop}%
\bibitem [{Note5()}]{Note5}%
  \BibitemOpen
  \bibinfo {note} {The complexity of preparing a quantum RAM device is beyond
  the scope of the present work.}\BibitemShut {Stop}%
\bibitem [{Note6()}]{Note6}%
  \BibitemOpen
  \bibinfo {note} {The presence of the term in brackets is necessary only if
  the system dealt with presents local interactions between the qubits. The
  nonlocal contributions results in a complexity aspect corresponding to $\log
  ^2(N)$, which is already taken into account in the terms proportional to
  $\log ^2(N)$ in the asymptotic notation.}\BibitemShut {Stop}%
\bibitem [{\citenamefont {Leveque}(2007)}]{Leveque2007}%
  \BibitemOpen
  \bibfield  {author} {\bibinfo {author} {\bibfnamefont {R.~J.}\ \bibnamefont
  {Leveque}},\ }\href@noop {} {\bibinfo {title} {Finite difference methods for
  ordinary and partial differential equations: steady-state and time-dependent
  problems}} (\bibinfo {year} {2007})\BibitemShut {NoStop}%
\bibitem [{\citenamefont {Chuang}\ and\ \citenamefont
  {Nielsen}(1997)}]{chuang1997}%
  \BibitemOpen
  \bibfield  {author} {\bibinfo {author} {\bibfnamefont {I.~L.}\ \bibnamefont
  {Chuang}}\ and\ \bibinfo {author} {\bibfnamefont {M.~A.}\ \bibnamefont
  {Nielsen}},\ }\bibfield  {title} {\bibinfo {title} {Prescription for
  experimental determination of the dynamics of a quantum black box},\ }\href
  {https://doi.org/10.1080/09500349708231894} {\bibfield  {journal} {\bibinfo
  {journal} {Journal of Modern Optics}\ }\textbf {\bibinfo {volume} {44}},\
  \bibinfo {pages} {2455} (\bibinfo {year} {1997})},\ \Eprint
  {https://arxiv.org/abs/https://www.tandfonline.com/doi/pdf/10.1080/09500349708231894}
  {https://www.tandfonline.com/doi/pdf/10.1080/09500349708231894} \BibitemShut
  {NoStop}%
\bibitem [{Note7()}]{Note7}%
  \BibitemOpen
  \bibinfo {note} {Non-Hermitian matrices are also tractable by the HHL
  algorithm, defining an operator $\protect \tilde {A}$, which is Hermitian and
  related to $A$.}\BibitemShut {Stop}%
\end{thebibliography}%

\end{document}